\documentclass[sigconf]{acmart}

\usepackage{subcaption} 
\usepackage{booktabs}
\usepackage{graphicx}%
\usepackage{multirow}%

\usepackage{amsmath,amssymb,amsfonts}%

\usepackage{mathrsfs}%
\usepackage[title]{appendix}%
\usepackage{xcolor}%
\usepackage{textcomp}%
\usepackage{manyfoot}%
\usepackage{siunitx}

\usepackage{listings}%
\usepackage{todonotes}

\usepackage{enumerate}
\usepackage[shortlabels]{enumitem}
\usepackage{algorithm,algcompatible}
\usepackage{csquotes}

\definecolor{boxgrey}{HTML}{F3F3F3}

\newcommand{\IGNORE}[1]{ }

\AtBeginDocument{%
  \providecommand\BibTeX{{%
    \normalfont B\kern-0.5em{\scshape i\kern-0.25em b}\kern-0.8em\TeX}}}

\definecolor{basicsaeblue}{HTML}{1E77B4}
\definecolor{topksaeorange}{HTML}{FF7F0E}
\definecolor{cosinetopksaegreen}{HTML}{2CA02C}

\copyrightyear{2026}
\acmYear{2026}
\setcopyright{none}
\begin{document}

\setlength{\abovedisplayskip}{0pt}
\setlength{\belowdisplayskip}{5pt}
\setlength{\abovedisplayshortskip}{0pt}
\setlength{\belowdisplayshortskip}{5pt}

\title{From Knots to Knobs: Towards Steerable Collaborative Filtering 
Using Sparse Autoencoders}

\author{Martin Spišák}
\email{martin.spisak@recombee.com}
\orcid{0009-0006-7763-5575}
\affiliation{%
  \institution{Recombee}
  \streetaddress{Vodi\v{c}kova 12/5, 120 00 Prague 2}
  \city{Prague}
  \country{Czech Republic}
  \postcode{120 00}
}
\additionalaffiliation{%
  \institution{Faculty of Mathematics and Physics, Charles University}
  \city{Prague}
  \country{Czech Republic}
}

\author{Ladislav Peška}
\email{ladislav.peska@matfyz.cuni.cz}
\orcid{0000-0001-8082-4509}
\affiliation{%
  \institution{Faculty of Mathematics and Physics, Charles University}
  \city{Prague}
  \country{Czech Republic}
}

\author{Petr Škoda}
\email{petr.skoda@matfyz.cuni.cz}
\orcid{0000-0002-2732-9370}
\affiliation{%
  \institution{Faculty of Mathematics and Physics, Charles University}
  \city{Prague}
  \country{Czech Republic}
}

\author{Vojtěch Vančura}
\email{vojtech.vancura@recombee.com}
\orcid{0000-0003-2638-9969}
\affiliation{%
  \institution{Recombee}
  \streetaddress{Vodi\v{c}kova 12/5, 120 00 Prague 2}
  \city{Prague}
  \country{Czech Republic}
  \postcode{120 00}
}
\additionalaffiliation{%
  \institution{Faculty of Mathematics and Physics, Charles University}
  \city{Prague}
  \country{Czech Republic}
}

\author{Rodrigo Alves}
\email{rodrigo.alves@fit.cvut.cz}
\orcid{0000-0001-7458-5281}
\affiliation{%
  \institution{Faculty of Information Technology, Czech Technical University}
  \city{Prague}
  \country{Czech Republic}
}

\renewcommand{\shortauthors}{Spišák, Peška, Škoda, and Alves}

\begin{abstract}
Sparse autoencoders (SAEs) have recently emerged as pivotal tools for introspection into large language models.
SAEs can uncover high-quality, interpretable features at different levels of granularity and enable targeted steering of the generation process by selectively activating specific neurons in their latent activations.
Our paper is the first to apply this approach to collaborative filtering, aiming to extract similarly interpretable features from representations learned purely from interaction signals. 

In particular, we focus on a widely adopted class of collaborative autoencoders (CFAEs) and augment them by inserting an SAE between their encoder and decoder networks. We demonstrate that such representation is largely monosemantic and propose suitable mapping functions between semantic concepts and individual neurons. We also evaluate a simple yet effective method that utilizes this representation to steer the recommendations in a desired direction.

\end{abstract}

\ccsdesc[500]{Information systems~Recommender systems}

\keywords{Sparse autoencoders, Interpretable recommender systems, Inprocessing recommendation adjustments}

\begin{teaserfigure}
\centering
  \includegraphics[width=0.9\textwidth]{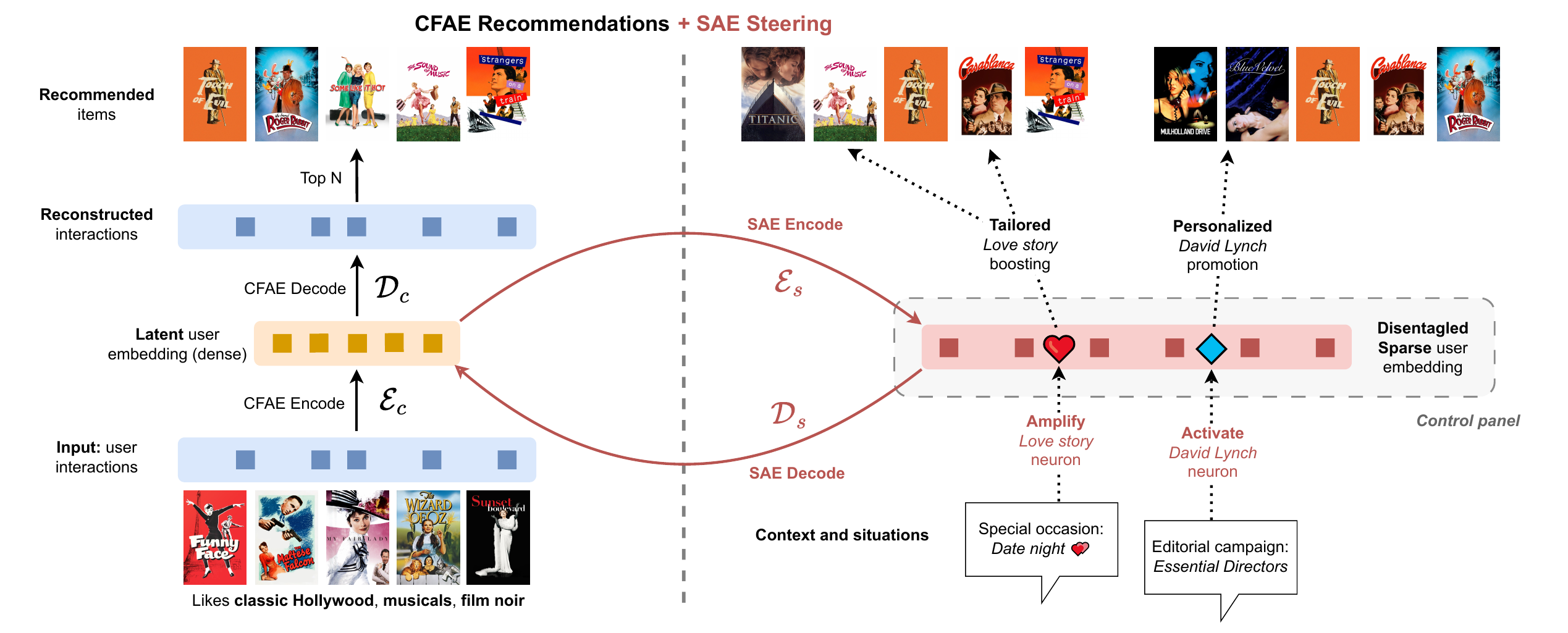}
  \caption{The proposed pipeline for steering CFAE-based recommendations. \normalfont{A sparse autoencoder (SAE) can disentangle user embeddings within collaborative filtering autoencoders (CFAEs), exposing a structured layer of interpretable “knobs.” Such knobs may be labeled based on their activation patterns and exposed to the users or editors in the form of a ``control panel''. They can steer the recommendations by gradually increasing the activation of such knobs that correspond to their current needs or desires.}}
  \label{fig:teaser}
\end{teaserfigure}

\maketitle

\section{Introduction}

\textbf{Collaborative-filtering} (CF) techniques have been dominating the recommender systems' research since the late 1990s, following their success in early benchmark datasets such as MovieLens and the Netflix Prize competition \cite{Koren2022}. While the dawn of deep learning partially shifted the focus towards approaches incorporating multiple input modalities \cite{10.1145/3640457.3688138}, pure CF techniques remain fundamental, especially for the initial stages of the recommendation pipeline, i.e., candidate retrieval. 

Among the plethora of CF algorithms, \textbf{collaborative autoencoders} (CFAEs)~\citep{sedhain2015autorec} belong to the most widely used approaches for this task. CFAEs
are recommendation methods designed to encode sparse user--item interaction data into a dense, lower-dimensional latent space and then reconstruct it, enabling the extraction of information crucial for matching users with items.  The encoder-decoder architecture can vary in complexity (from shallow~\citep{10.1109/ICDM.2011.134,10.1145/3308558.3313710} to deep structures~\citep{li2015deep,liu2018novel}, including variational approaches~\citep{10.1145/3178876.3186150, 10.1145/3336191.3371831}), allowing the model to capture different levels of abstraction and adapt to the specific characteristics of the data.
By focusing on the most relevant features of user behavior and item characteristics during reconstruction, these models reveal latent patterns and hidden relationships that are not immediately apparent in the raw data. This versatility, combined with strong recommendation performance, has made them a popular choice among researchers and practitioners for addressing challenges such as data sparsity and noise~\citep{li2024survey,vanvcura2023scalable}, making these models a surging trend in the recommendation industry~\cite{10.1145/3308558.3313710,jeunen2022embarrassingly,vanvcura2024beeformer,sansa}.

However, despite their flexibility and robust performance, understanding the underlying mechanisms and characteristics hidden in the latent structure of CFAEs remains a very challenging task~\citep{xu2019explainable,longo2020explainable,abusitta2024survey}. The abstract encoder-decoder mechanism not only results in opaque model behavior, making it difficult to track how recommendations are derived, but also limits the capacity of different agents (e.g., users, editors, providers) to influence or steer the recommendation process. This lack of transparency is inherent also for other CF techniques based on latent representations, and poses significant challenges when diagnosing issues, fine-tuning performance, ensuring fairness, and granting meaningful control over recommendation outcomes. Even shallow architectures, which might appear straightforward at first glance, can exhibit unexpected behaviors on complex real-world datasets~\citep{absease}. This complicates efforts to trace decision paths and adjust parameters to satisfy both business and ethical standards~\citep{10.1093/idpl/ipx022}.

Understanding internal mechanisms and latent model structures is a broader challenge, not limited to CF, and has been explored across many areas of machine learning.
For instance, the field of mechanistic interpretability has recently made substantial progress in tackling this problem, driven by rapid advancements in large language models (LLMs)~\citep{bereska2024mechanisticinterpretabilityaisafety}.
One promising approach involves using a different type of autoencoder -- a \textbf{sparse autoencoder (SAE)} --  to identify highly interpretable, monosemantic features (i.e., a dictionary of concepts)\footnote{We understand the term \textit{concept} as a latent yet meaningful property of an item.} in the residual streams of language models~\citep{cunningham2023sparseautoencodershighlyinterpretable, bricken2023towards}. This is achieved by disentangling typically polysemantic activation neurons through high-dimensional, sparsely activated latent representations. The disentangling projection is trained in a self-supervised manner using sparsity-inducing activation functions.
Promisingly, recent work has successfully applied this approach to highly expressive models, including state-of-the-art LLMs~\citep{gao2024scalingevaluatingsparseautoencoders, templeton2024scaling}.

In this research, we investigate the use of \textbf{SAEs} as an interface for interpreting and steering the outputs of collaborative filtering algorithms, specifically \textbf{CFAEs}. We illustrate this approach in Figure~\ref{fig:teaser}. On the left, a standard CFAE encodes a user’s (sparse) interaction vector into a dense latent embedding, which is then decoded, producing recommendation scores for all candidate items and generating Top-$N$ recommendations. 
This process alone offers limited options for external influence, apart from modifying the input interaction vector directly. 
Therefore, as shown on the right of Figure~\ref{fig:teaser}, we propose inserting an SAE between the CFAE’s encoder and decoder. 
This SAE ``hook'' maps the dense latent user representation to a higher-dimensional, sparsely activated user code -- effectively exposing a control panel of interpretable knobs (neurons), each corresponding to a distinct concept. For example, a neuron may be associated with \emph{love stories} (depicted as a red heart), or \emph{David Lynch} (blue diamond).\footnote{Interestingly, our experiments (Section~\ref{sec:intersae}) show that meaningful neuron--concept correspondence emerges through self-supervised SAE training on purely interaction-based data -- that is, even without access to item metadata.}
To steer the recommendations, we may amplify one or more of the knobs (e.g., boost the \emph{love story} neuron in a \emph{date night} context). The modified sparse code then passes through the SAE decoder and subsequently the CFAE decoder to generate recommendations \emph{steered} toward the desired theme while maintaining original user preferences as much as possible. By exposing and manipulating these monosemantic controls, our method enables fine-grained, human-understandable steering of recommendations. 
Moreover, different agents can initiate these semantic boosts: for example, (1) a user might increase the \emph{love story} knob for a romantic evening via a graphical interface, while (2) an editor could selectively boost the \emph{David Lynch} knob (even if it was originally inactive and not part of the user’s initial interests) to craft curated, editorial recommendations. 
Ultimately, our approach presents a key interface for seamless, personalized control over recommendations through semantic-level interventions, rather than by manual item selection.

Although previous work has addressed the steering of general recommendation models~\citep{10.1145/2365952.2365964,10.1145/2792838.2800179,10.1145/2365952.2365966} as well as interpretability of CFAE-like models~\citep{epstein2019generalization,absease,10.1145/3523227.3551482,refinetti2022dynamics,guo2024dualvae,vsafavrik2022repsys}, 
none of these approaches has applied SAEs to generate interpretable user embeddings while \emph{simultaneously} steering recommendations.
Moreover, no previous study has investigated the nuanced characteristics of user--item interaction reconstruction quality in CFAEs, nor has any work systematically examined the sparse dictionary of SAEs in relation to structured metadata for recommender systems. To address these research gaps, our main contributions are listed as follows:


\begin{itemize}
    \item We propose a pipeline for steerable recommendations based on CFAEs. In the pipeline, we first (i) train a selected CFAE as usual, and (ii) input an SAE ``hook'' to learn a sparse representation of the CFAE embeddings. Then (iii), based on common item-neuron activation patterns, we assign labels for individual neurons, and finally (iv) disclose the resulting ``control panel'' for steering purposes.


    \item We show that the SAE's sparse reconstruction can preserve downstream accuracy of the unmodified CFAE. 


    \item We show that many SAE neurons correspond to interpretable concepts - even when no metadata was used during the training.

    \item We demonstrate the utility of SAEs for targeted steering via concept-specific neuron activation.
    


\end{itemize}

\section{Background and Related Work}
\paragraph{\textbf{Basic Notation}} For a set of users $\mathcal{U} = \{1, 2, \ldots, m\}$ and a set of items $\mathcal{I} = \{1, 2, \ldots, n\}$, let $X \in \{0,1\}^{m \times n}$ be a partially-observed matrix of binary interactions (e.g., clicks, purchases), where the rows represent users and the columns represent items, i.e. $x_{ij}$ denotes the presence or absence of an interaction between user-$i$ and item-$j$. Typically, the matrix $X$ is very sparse, meaning only a small fraction of user--item interactions have been observed. Moreover, let $x_i \in \{0,1\}^n$ denote the $i$-th row vector of the matrix $X$. Using this notation, the goal of a recommender system based on collaborative filtering is to learn a function $f_\theta:\mathcal{U}\times\mathcal{I}\to\mathbb{R}$ (parametrized by $\theta$) that predicts a relevance score $r(u,i) = f_\theta(u,i,X)$ for user $u \in \mathcal{U}$ and item $i \in \mathcal{I}$ using the interaction matrix $X$.

Our methodology employs an autoencoder architecture.
A core optimization problem for an autoencoder can be defined as

\begin{align}
& \min_{\mathcal{E}, \mathcal{D}}\  \sum_{i=1}^{m} \ell\bigl(x_i, \sigma(x_i)\bigr) \nonumber
&\text{with} \ \ \sigma(x) = \bigl(\mathcal{D} \circ \mathcal{E}\bigr)(x),
\end{align}

\noindent where $\mathcal{E}$ and $\mathcal{D}$ denote, respectively, the encoder and decoder functions, $f \circ g$ denotes the composition of functions $f$ and $g$, and $\ell(x,y)$ is a reconstruction loss that measures the discrepancy between vectors $x$ and $y$.
Since we embed SAE within the CFAE in the proposed pipeline, the nested architecture can be described as

\begin{align}
 \sigma_{c,s}(x) = \bigl(\mathcal{D}_c \circ \mathcal{D}_s \circ \mathcal{E}_s \circ \mathcal{E}_c\bigr)(x),
\end{align}
where $\mathcal{E}_c$ and $\mathcal{D}_c$ represent encoder and decoder of the ``outer'' CFAE, while $\mathcal{E}_s$ and $\mathcal{D}_s$ represent the ``inner'' SAE.

\paragraph{\textbf{Collaborative Autoencoders}}
CFAEs are, together with neighborhood-based \cite{Nikolakopoulos2022}, graph-neural \cite{10.1145/3568022}, or two-tower models \cite{10.1145/2959100.2959190}, among the most popular choices in production recommender systems. CFAEs are especially valued for the initial retrieval phase of the recommendation generation due to their efficient representation learning under sparse implicit feedback and high scalability, making them suitable for large-scale retrieval pipelines~\cite{vanvcura2025evaluating}. In our work, we focus on two popular and complementary classes of CFAEs: linear and variational autoencoders.

\textit{Linear autoencoders}~\cite{10.1109/ICDM.2011.134, 10.1145/3308558.3313710, 10.1145/3523227.3551482, sansa, 10.5555/3495724.3497368} are \emph{shallow} models favored for their straightforward, easy-to-implement, and generally scalable design. Probably the simplest and most popular representative is the EASE algorithm~\cite{10.1145/3308558.3313710}.
However, its $O(n^2)$ model size limits scalability in terms of item quantity~\cite{10.1145/3523227.3551482, sansa, 10.5555/3454287.3454778}, while with a fused encoder and decoder, there is no direct function to construct intermediate representations.
Therefore, instead of EASE, we utilized its low-rank approximation ELSA~\cite{10.1145/3523227.3551482}, which effectively mitigates both issues while retaining a highly competitive performance.

\textit{Variational autoencoders} (VAE)~\citep{10.1145/3178876.3186150, 10.1145/3336191.3371831, 10.1145/3663364, 10.1145/3298689.3347015} belong to a family of \emph{deep, generative} models and are known for their ability to model complex data distributions through probabilistic latent representations, offering enhanced flexibility and expressiveness compared to linear models. Unlike classical autoencoders that project inputs to a single point in the bottleneck latent space, VAE representations $z_i = \mathcal{E}_c(x_i)$ are transformed via a function to produce a probability $\pi(y)$ distribution over the items, from which the click history $x_i$ is assumed to have been generated. In this study, we experimented with MultVAE~\citep{10.1145/3178876.3186150}, one of the best-known VAE representatives.

\paragraph{\textbf{Sparse Autoencoders}}
Sparse autoencoders (SAEs)~\cite{bricken2023towards,makhzani2014ksparseautoencoders} are neural network architectures trained to reconstruct their input while promoting sparsity in high-dimensional\footnote{The hidden layer dimension is often several times higher than the input dimension.} latent activations. Unlike traditional autoencoders, which compress input into a lower-dimensional bottleneck layer, SAEs map their input to a more expressive latent space but constrain the activations by allowing only a few nonzero entries. There are several ways to impose such an objective, e.g., by regularizing the size of the learned latent representation ~\cite{bricken2023towards}, or by imposing a rigid constraint on the number of active latents (using, e.g., Top-$k$~\cite{makhzani2014ksparseautoencoders} or batch Top-$k$~\cite{bussmann2024batchtopksparseautoencoders} activation functions). The resulting representations often tend to be more interpretable, with individual neurons representing distinct, monosemantic concepts. Promisingly, recent works have demonstrated the effectiveness of SAEs in disentangling representations from highly expressive models, including state-of-the-art LLMs~\cite{gao2024scalingevaluatingsparseautoencoders, templeton2024scaling}.


\paragraph{\textbf{Control and Transparency in Recommender Systems}}
While the prevalent recommendation paradigm is to supply users with recommendations directly with no user intervention, several works suggest that users benefit (e.g., through better recommendation acceptance) when they have some degree of control over the system \cite{10.1145/1357054.1357222, 10.1145/2365952.2365964, 10.1145/2792838.2800179, 10.1145/2365952.2365966, 10.1080/10447318.2023.2262796,PARRA201543,10.1145/3523227.3546772}.
For example, Parra and Brusilovsky \cite{PARRA201543} proposed an interactive interface that allowed users to specify weights for fusing individual recommending methods. Liang and Willemsen \cite{10.1145/3523227.3546772} let users control the trade-off between personalized and representative music recommendations while exploring a new genre. Similarly, Millecamp et al. \cite{10.1145/3209219.3209223} developed a UI for adjusting parameters like danceability for music recommenders. Nevertheless, these approaches relied on post-processing to incorporate user control, while our proposal is capable of addressing it within the model itself.
Closest to our work is the TEASER model proposed by De Pauw et al.~\cite{depauw2022modellingusersitemmetadata}. TEASER is a hybrid interactive recommendation model based on EDLAE~\cite{10.5555/3495724.3497368},
a linear CFAE similar to one of our backbones.
Nevertheless, TEASER is trained via a combination of user interactions and item metadata and explicitly defines latent features corresponding to the metadata. These features are then used as ``knobs'' for recommendation steering.
In contrast, our ``knobs'' emerge organically within an SAE embedded into a CFAE and reflect solely the properties implicitly available in the interaction data alone; we use metadata merely to label the knobs.

While user-side control over recommendations has already received notable attention from the research community, editorial-side control is also critically important in industry settings, despite being less explored to date \cite{10.1145/3687151.3687152,10.1145/3340631.3394864}. Existing approaches typically rely on simple item-level boosting or intervene during post-processing of recommendations \cite{10.1145/3477495.3531890,10.1145/3340631.3394864}, which limits the flexibility of intervention due to dependence on the initial item rankings. Instead, our approach paves the way for fine-grained in-process intervention already at the candidate generation stage.

The labeled SAE neurons learned by our method offer a promising solution to the interpretability challenges of CFAEs. Specifically, the neurons activated by a user's historical feedback can be directly interpreted as the model's internal representation of the user's preferences. Furthermore, the mapping from labeled neurons to item features provides a way to explain how specific items align with these inferred preferences. 
Crucially, unlike most existing methods that focus on post hoc justifications for recommended items~\cite{guesmi2023justificationvstransparencyvisual,park2020jrecsprincipledscalablerecommendation,Balog_2023,muhammad2016use,4648950,Mauro2023}, our technique, similar to~\cite{epstein2019generalization,wang2024interpretinternalstatesrecommendation}, enables inspection of the model’s internal mechanics and thus contributes to the transparency of the recommendations as well.

\paragraph{\textbf{Using SAEs in CF techniques}} We are not aware of any directly related approaches aiming to use SAEs to enable steering of CFAE-based recommendations. While some proposals for using metadata for CFAE interpretation exist~\cite{depauw2022modellingusersitemmetadata}, none have used SAEs for that purpose. 
In fact, to the best of our knowledge, Wang et. al. \cite{wang2024interpretinternalstatesrecommendation} is the only representative other work that uses SAEs to interpret the recommendation models.
Nevertheless, they focused on a different recommendation task (sequential recommendation) and 
and evaluated their method exclusively on a single architecture, SASRec~\cite{8594844}, a transformer-based model similar to those typically studied in prior SAE work.
Moreover, while Wang et al. briefly mention the possibility of recommendation steering, this aspect is not systematically explored or evaluated.
In contrast, our work focuses on establishing the foundations for recommendation steering by using SAEs within CFAEs. We demonstrate that not all CFAE architectures are equally robust to sparse reconstruction and that the choice of reconstruction loss can have positive or detrimental effects. This corrects the over-optimistic conclusions about SAE incorporation into an arbitrary recommending algorithm \cite{wang2024interpretinternalstatesrecommendation}. 


\section{Methodology}
\label{sec:autoencoders_descr}
To enable the steering of CFAE-based recommenders, we propose a four-stage pipeline as follows: First, the CFAE model is trained as usual. Then, we nest an ``SAE hook'' aiming to reconstruct the dense CFAE embeddings using a wider yet sparsely activated layer. Next, we create a mapping between available semantic concepts and individual neurons on the SAE's hidden layer. Finally, the labeled neurons are supplied as a ``control board'' to  interested parties, enabling targeted in-process steering of  recommendations by adjusting the neurons' activation patterns.

Note that, in principle, each stage is independent of the composition of the others. This makes the pipeline highly modular, where individual variants of each stage can be deployed in a plug-and-play fashion. Therefore, in this foundational work, we intentionally focused on evaluating relatively simple variants for each stage, thereby providing a straightforward, scalable, highly reproducible, yet competitive baseline for future research. 

\subsection{Network Architecture}
\label{sec:sae}
The network architecture overview is depicted in Figure \ref{fig:teaser}. In particular, we propose a nested autoencoder architecture, where a CFAE serves as the outer layer, while an SAE serves as the inner layer. The resulting network can be seen as a function  $\sigma_{c,s}(x) = \bigl(\mathcal{D}_c \circ \mathcal{D}_s \circ \mathcal{E}_s \circ \mathcal{E}_c\bigr)(x)$, where $x$ represent a vector of user's interactions. Naturally, while in general $\mathcal{E}_c: \mathbb{R}^n \rightarrow \mathbb{R}^r$ and $\mathcal{E}_s: \mathbb{R}^p \rightarrow \mathbb{R}^d$, we implicitly assume $p \equiv r$ when we nest an SAE within a CFAE.

\subsubsection*{\textbf{CFAEs}} As representatives for the CFAE, we selected well-known ELSA~\cite{10.1145/3523227.3551482} and MultVAE~\cite{10.1145/3178876.3186150} algorithms. Note that in the case of ELSA, which optimizes
$\min_{A} \|X - X(AA^\top - I)\|^2_{F}  \quad \text{s.t.} \quad \|a_i\|_2 = 1$, we may consider its first step as encoder, i.e., $\mathcal{E}_c(x) = A^Tx$, while the remainder as decoder, i.e., $(\mathcal{D}_c \circ \mathcal{E}_c)(x) = A\mathcal{E}_c(x) - x$. For MultVAE, the embeddings are taken from the mean head of the encoder network.

\subsubsection*{\textbf{SAEs}}As representatives for SAE, we experimented with a basic ReLU SAE~\cite{bricken2023towards} (which, for the sake of clarity, we refer to as \emph{Basic SAE}) and a $k$-sparse autoencoder, \emph{TopK SAE}~\cite{makhzani2014ksparseautoencoders}. 

\paragraph{Basic SAE}
For an input vector $y_i \in \mathbb{R}^{p}$, \cite{bricken2023towards} proposes a \newline$d$-dimensional ReLU SAE encoder $\mathcal{E}_s$ and decoder $\mathcal{D}_s$ defined as:

\begin{align*}
\mathcal{E}_s(y_i) &= \mathrm{ReLU}\bigl(W_{\mathcal{E}_s}(y_i - b_{\mathcal{D}_s}) + b_{\mathcal{E}_s}\bigr) \text{ and }\\
\mathcal{D}_s(\mathcal{E}_s(y_i)) &= W_{\mathcal{D}_s}\mathcal{E}_s(y_i) + b_{\mathcal{D}_s},
\end{align*}

\noindent with learnable parameters $W_{\mathcal{E}_s} \in \mathbb{R}^{d \times p}$, $b_{\mathcal{E}_s} \in \mathbb{R}^{d}$, $W_{\mathcal{D}_s} \in \mathbb{R}^{p \times d}$, and $b_{\mathcal{D}_s} \in \mathbb{R}^{p}$. During both training and inference, input vectors are standardized before being passed through the SAE and de-standardized on the output.

\paragraph{TopK SAE} 
The $k$-sparse autoencoder described in \cite{makhzani2014ksparseautoencoders} differs from the Basic SAE only in the encoder, which is defined as:

\begin{equation*}
    \mathcal{E}_s(y_i) = \mathrm{TopK}\bigl(W_{\mathcal{E_s}}(y_i - b_{\mathcal{D_s}}), k\bigr),
\end{equation*}

\noindent where $\mathrm{TopK}(z,k)$ is an activation function that zeros out all but the $k \in \mathbb{N}$ positive largest values of vector $z$; $k$ is a hyperparameter that controls the activation density. Following~\cite{gao2024scalingevaluatingsparseautoencoders}, we include a ReLU activation before applying the TopK function, forcing all activations to be non-negative. To align the two architectures, we incorporate an encoder bias term (similar to Basic SAE), resulting in:

\begin{equation*}
     \mathcal{E}_s(y_i) = \mathrm{TopK}\Bigl(\mathrm{ReLU}\bigl(W_{\mathcal{E_s}}(y_i - b_{\mathcal{D_s}}) + b_{\mathcal{E_s}}\bigr), k\Bigr).
\end{equation*}

\noindent This formulation makes the model equivalent to Basic SAE, with the only difference being the addition of a TopK activation.

\paragraph{Optimization Objective} 
To optimize our SAE variants, we use the following loss function:

\begin{equation*}\label{sae_loss}
    \min_{\mathcal{E}_s,\mathcal{D}_s} \sum_{i=1}^{m} \ell_s(y_i) = \sum_{i=1}^{m}  \|y_i - \bigl(\mathcal{D}_s \circ \mathcal{E}_s\bigr)(y_i)\|_2^2 + \lambda \|\mathcal{E}_s(y_i)\|_1,
\end{equation*}

\noindent where $\lambda$ is an $\mathrm{L}_1$-norm regularization hyperparameter. Note that while $\lambda$ serves as the primary controlling parameter for sparsity in Basic SAE, in TopK SAE it mainly prevents large spikes in activations and plays only a secondary role in enforcing sparsity. For the sake of simplicity, and because the SAEs we train are not very wide,
we do not incorporate auxiliary losses, as described in~\cite{gao2024scalingevaluatingsparseautoencoders}.

\subsubsection*{\textbf{Network training}} We propose to train the architecture in two steps. First, train only the CFAE part to reconstruct the hidden part of user-item interactions. Second, train SAEs to reconstruct user embeddings derived from interaction data using a pretrained CFAE. This simulates the real-world modularity, where CFAE models are readily available in the recommendation pipelines, and we only need to augment them with the SAE hook.



\subsection{Concept-Neuron Mapping}
\label{sec:concept_neuron_mapping}
The next step toward SAE-based steering is to determine whether a correspondence has emerged between sparse neurons and semantic concepts -- that is, to identify which ``knobs'' can be manipulated to steer recommendations in a desired way. This requires evidence of a reasonably narrow mapping between concepts and neurons in both directions. 

To label the ``knobs'', we employ a simple yet effective method that leverages the available item metadata, e.g., user-created textual tags as collected for the ML-25M dataset. Let $\mathcal{T}$ denote the set of all such tags.
We first estimate the joint tag-item distribution, $\hat{p}(t \in \mathcal{T}, i \in \mathcal{I})$, by normalizing the sparse tag assignment count matrix.\footnote{In different contexts, one may utilize any other prior information specifying how relevant a particular tag is for the item at hand.} 
Next, to link SAE neurons with entities in the data, we compute sparse representations for every item by passing its one-hot encoding through the encoder part of the network, i.e., $z_i = \bigr(\mathcal{E}_s \circ \mathcal{E}_c\bigr)(\texttt{onehot}(i))$.
Multiplying the empirical tag-item distribution matrix by the item-neuron sparse activations produces a tag--activation matrix $\mathcal{M}: \mathbb{R}^{|\mathcal{T}|\times d}$, where $d$ is the dimensionality of the SAEs' hidden layer. 

From the $\mathcal{M}$ matrix, we derive TF-IDF scores in two orientations: $M_{t \rightarrow n}$, treats tags as terms and neurons as documents, while $M_{n \rightarrow t}$ treats neurons as terms and tags as documents. Note that this duality carries important semantical nuances. In particular, $\text{argmax}_n M_{t \rightarrow n}$ identifies, for each tag, the neuron whose firing is \textbf{most unique} to that tag, while $\text{argmax}_t M_{t \rightarrow n}$ selects, for each neuron, the tag that \textbf{best characterizes} its overall activity. Conversely, $\text{argmax}_t M_{n \rightarrow t}$ finds, for each neuron, the tag that elicits its \textbf{most distinctive} response, and $\text{argmax}_n M_{n \rightarrow t}$ finds, for each tag, the neuron that \textbf{most representatively} encodes it.

\subsection{Recommendation Steering}
\label{sec:steering}
Finally, the established concept--neuron mappings can be made available to interested parties as a means of altering the recommendations to a desired direction. Nevertheless, before describing the exact mechanism, let us first formalize our understanding of this steering process. 
\medskip


\noindent \textbf{Definition: Segment} $\mathcal{S} \subseteq \mathcal{I}$ is a subset of items that share a common \textit{concept}, formally represented by a binary indicator function
$s(i) = 1$ if $i \in S$, and $s(i) = 0$ otherwise. 

\noindent \textbf{Definition: Steering toward a target segment $\mathcal{S}$} is a controlled transformation of the user–item relevance function $f_\theta$ such that the predicted relevance is adjusted to favor items from the target segment $\mathcal{S}$ while maintaining consistency with the original user preference model. Formally, steering induces a new relevance estimator $f_{\theta, \mathcal{S}, \alpha}(u, i) = \Phi\big(f_\theta(u, i), g(i, \mathcal{S}), \alpha\big)$, where $g(i, \mathcal{S})$ quantifies the relevance alignment of item $i$ to the segment $\mathcal{S}$, and $\Phi(\cdot)$ is a steering operator which combines the base relevance and the steering alignment, controlled by intensity parameter $\alpha$.
\medskip

To realize the steering in the nested autoencoder architecture, we start with a sparse vector $z_u = \bigl(\mathcal{E}_s \circ \mathcal{E}_c\bigr)(x_u)$, representing a user 
$x_u$  and a specific neuron $j$ to be boosted (determined via some concept-neuron mapping).\footnote{The extensions to simultaneous boosting or suppression of multiple neurons / multiple concepts are trivial, so we omit them for the sake of space.} We first normalize $z_u$ to unit sum, obtaining $\bar{z}_u$, and then construct a steered representation by a convex combination of the original profile and the targeted concept emphasis: 
$\tilde{z}_u = (1 - \alpha) \times \bar{z}_u + \alpha \times \texttt{onehot}(j)$ for $\alpha \in \left[0,1\right]$. The steered representation $\tilde{z}_u$ is then passed through the decoder part of the network to obtain the recommendations. 

\section{Experiments}
We systematically establish the foundations for SAE-enabled 
steering of CFAEs.
First, we investigate the robustness of CFAE latent representations to disentanglement and sparse reconstruction, analyze the behavior of different SAEs, and explore the attainable trade-off between the sparsity of SAE neurons and its negative impact on downstream recommendation quality. 
Second, we study the ``knobs'' emerging during self-supervised, interaction-only training and search for concepts that these neurons might represent.
Finally, we demonstrate how the discovered concept-neuron mapping can be leveraged to 
steer recommendation outcomes. The source codes for reproduction purposes, as well as additional results, are available at \url{https://anonymous.4open.science/r/knots-to-knobs}.

\begin{table}[t]
\centering
\caption{Dataset statistics.}
\label{tab:dataset_stats}
\begin{tabular}{l c c c c }
\toprule
Dataset   & \#Users & \#Items & \#Interactions & Density (\%) \\
\midrule
ML-25M  & 160 776  & 40 857  & 12.4M & 0.19\%  \\
MSD & 571 355  & 41 140   & 33.6M         & 0.14\% \\
\bottomrule
\end{tabular}
\end{table}

\subsection{Preliminaries}
\paragraph{Datasets} We conduct experiments on two well-known recommendation datasets, namely MovieLens 25M (ML-25M)~\cite{10.1145/2827872} and Million Song Dataset (MSD)~\cite{BertinMahieux2011TheMS}, and follow the same pre-processing steps as described in \cite{10.1145/3178876.3186150}:
\begin{itemize}
    \item For ML-25M, we first convert the explicit ratings into binary implicit feedback by treating ratings $\geq 4$ as positive interactions (with a value of 1) and discarding the rest. We then remove users with fewer than 5 interactions.
    \item For MSD, we binarize play counts and consider them as implicit feedback. Next, we remove items with fewer than 200 interactions and users with fewer than 20 interactions.
\end{itemize}
We split each dataset into training, validation, and test sets, employing the principle of strong generalization -- ensuring that users are disjoint between the splits. We reserve 10\% of users for the test set and an additional 10\% for validation, leaving 128,621 and 457,084 users for training on ML-25M and MSD, respectively. Dataset statistics are summarized in Table \ref{tab:dataset_stats}.

\paragraph{Training and Evaluating CFAEs}
For each dataset, we train ELSA and MultVAE in three sizes: "small", "medium", and "large". In particular, for ELSA, we use embedding dimensions $d \in \{512, 1024, 2048\}$. For MultVAE, we select a neural architecture $[d_{\text{input}} \rightarrow 3d \rightarrow d]$, where $d_{\text{input}}$ is the number of items and $d \in \{256,512, 1024\}$, with a $\tanh$ activation\footnote{The final layer has two heads -- one for the mean and one for the variance. No activation is applied to these heads.}. The decoder mirrors the encoder structure and applies a softmax activation to the output. This architecture largely follows the suggestions from the original paper \cite{10.1145/3178876.3186150}. 

To contextualize the results discussed in the next subsection, we evaluate the CFAEs in terms of Recall@20 and nDCG@20 on the held-out sets. 
The evaluation targets are randomly selected as 20\% of each user's interactions, with the remainder serving as inference input. 
Although the peak performance of CFAEs is not the primary focus of our experiments, the results for ELSA are on par with those reported in \cite{10.1145/3523227.3551482} (Recall@20 of 0.390 - 0.397 for ML-25M and 0.245 - 0.298 for MSD; full details are available from the appendix). For MultVAE, our results (Recall@20 of 0.375 - 0.383 for ML-25M and 0.211 - 0.252 for MSD) were approximately 5\% lower than those of \cite{10.1145/3178876.3186150}, which we attribute to a simpler annealing strategy and a less extensive hyperparameter search.

\paragraph{Training SAEs}
We trained 12 variants of Basic SAE and 12 variants of TopK SAE for every combination of dataset, CFAE model, and its size, resulting in 288 sparse autoencoders in total.
For both SAE architectures, we varied the width-to-input dimension ratios $\{2, 4, 8\}$, while for Basic SAE, we tuned the $\mathrm{L}_1$ coefficient, and for TopK SAE, we tuned the $k$ hyperparameter. Each SAE was trained on the embeddings of the training user interactions generated by the encoder $\mathcal{E}_c$ of its parent CFAE model.


To put the scale of our training into perspective, our runs process 128,621 and 457,084 unique user embeddings for ML-25M and MSD, respectively, and train SAEs for 250 epochs, totaling approximately 32 million and 114 million processed tokens per dataset. In contrast, for example, \citet{bussmann2024batchtopksparseautoencoders} trains TopK SAEs on 1 billion token activations from LLMs, making their training approximately 10–30$\times$ larger in terms of training volume.

\begin{figure*}[tb]
    \centering
    \begin{subfigure}{0.32\textwidth}
        \centering
        \includegraphics[width=\linewidth]{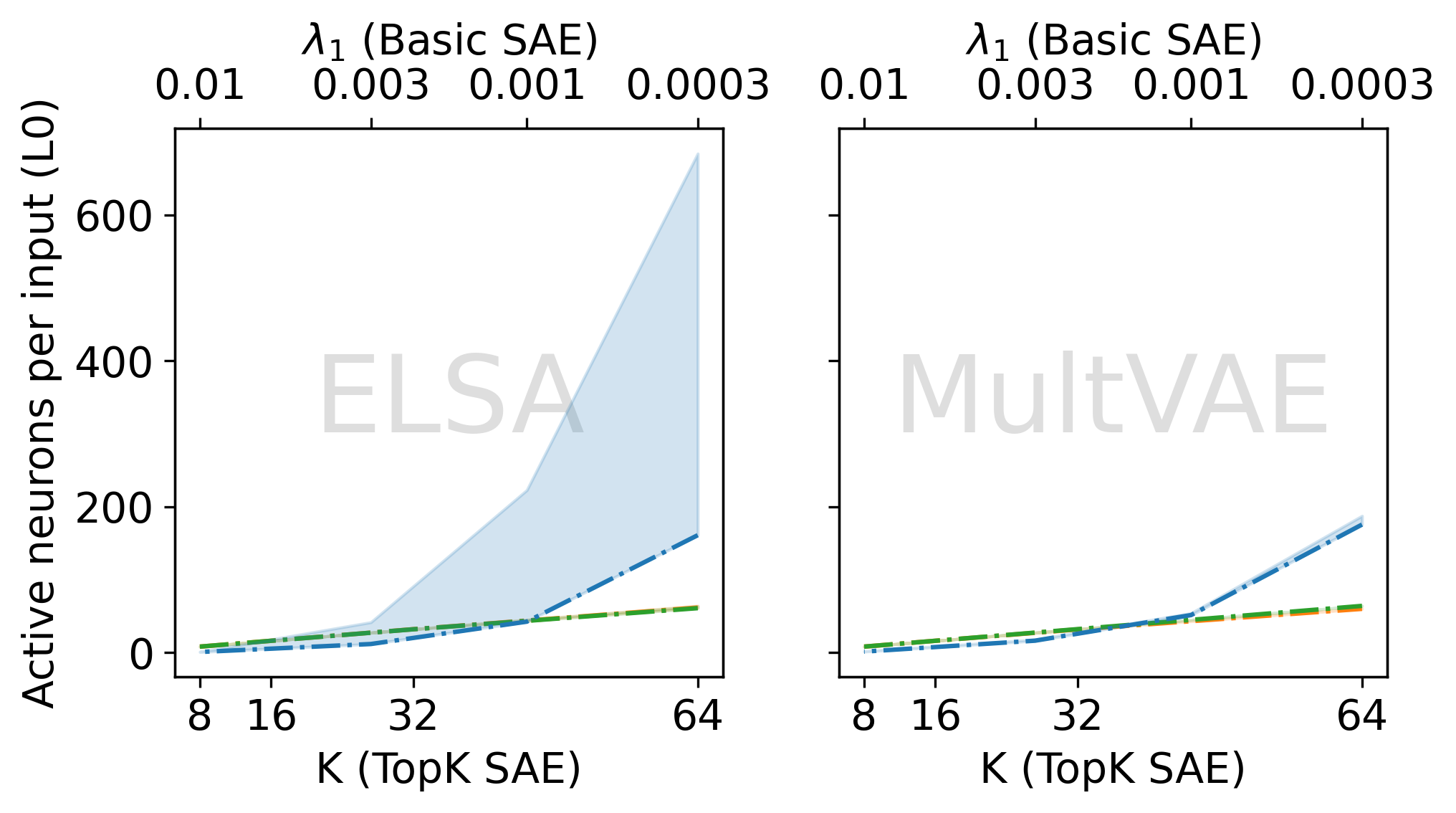}
        \caption{Activation density}
        \label{fig2:sub1}
    \end{subfigure}
    \hfill
    \begin{subfigure}{0.32\textwidth}
        \centering
        \includegraphics[width=\linewidth]{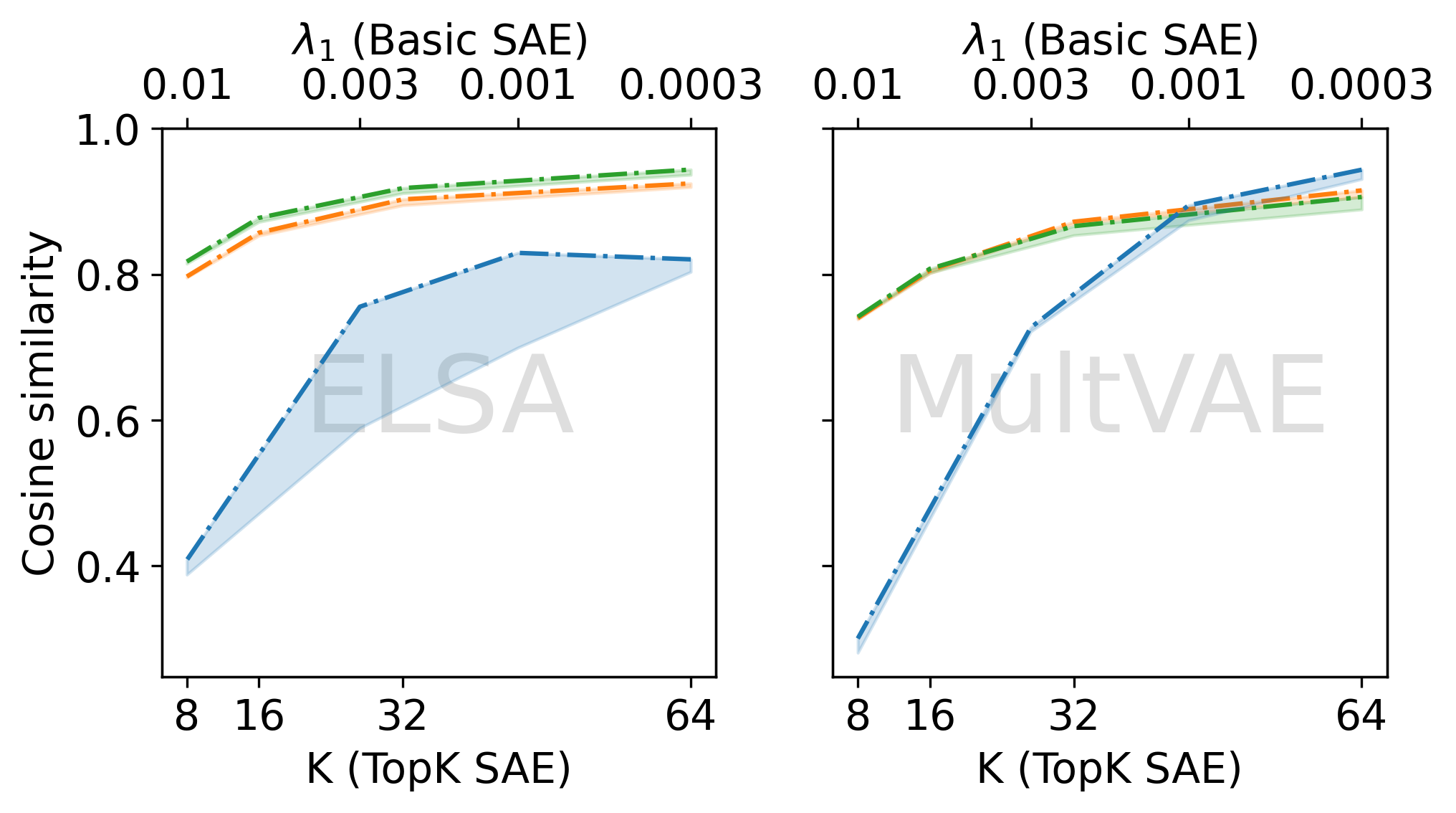}
        \caption{Embedding reconstruction accuracy}
        \label{fig2:sub2}
    \end{subfigure}
    \hfill
    \begin{subfigure}{0.32\textwidth}
        \centering
        \includegraphics[width=\linewidth]{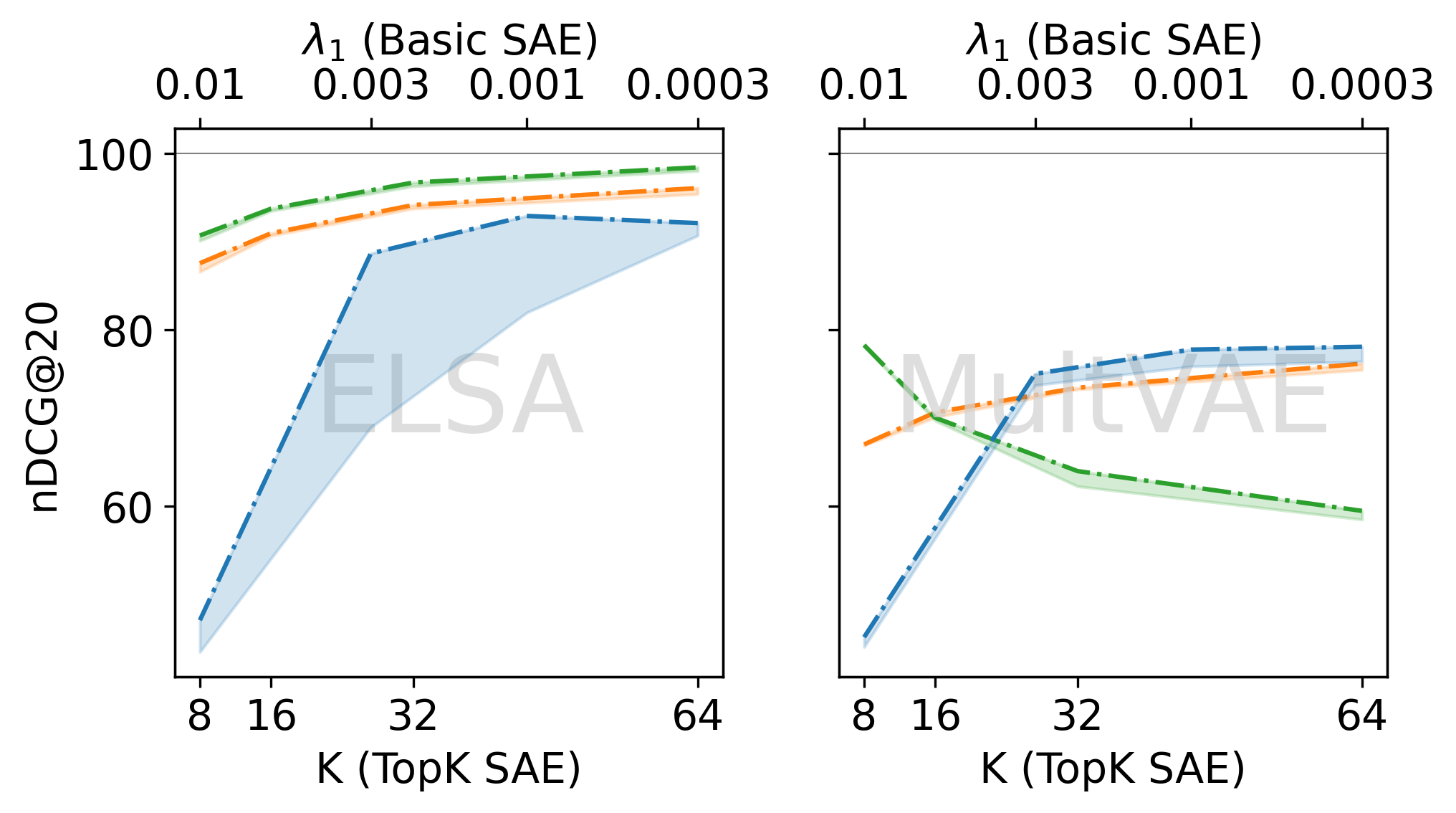}
        \caption{Recovered downstream performance in \%}
        \label{fig2:sub3}
    \end{subfigure}
    \caption{
    Effects of SAE reconstruction. 
    \normalfont{
    Results show SAEs trained on 1024-dimensional CFAE embeddings with various sparsity-inducing hyperparameters ($k, \ \lambda_1$).
    (1) \textcolor{topksaeorange}{\textbf{TopK SAE} (orange)} 
    is simple to use and achieves a superior sparsity--accuracy trade-off 
    compared to \textcolor{basicsaeblue}{\textbf{Basic SAE} (blue)}, 
    which is very sensitive to hyperparameter selection. (2) \underline{TopK SAEs reconstruct cosine-like embeddings (ELSA) accurately} \underline{and with minimal performance degradation.} Variational embeddings (MultVAE) are difficult to reconstruct without sacrificing performance.\newline
    \textbf{Ablation:} Replacing the $\mathrm{L}_2$ reconstruction loss in TopK SAE 
    with \textcolor{cosinetopksaegreen}{\textbf{cosine similarity loss} (green)} 
    further improves the sparsity-accuracy trade-off for cosine-like embeddings (ELSA). However, this change no longer guarantees a small Euclidean distance between variational embeddings and their reconstructions, breaking the downstream performance in the case of MultVAE.} }
    \label{fig2:sparsity_vs_accuracy}
\end{figure*}

\subsection{Reliability of Sparse Reconstruction}\label{sec:3_2}




Before applying sparse neurons for steering, we first address a foundational question: \emph{how robust and accurate is sparse reconstruction of CFAE user embeddings?} This experiment establishes whether SAE-based reconstruction can be reliably used at inference time without degrading downstream recommendation quality. If sparse reconstruction is unstable or lossy, it risks introducing harmful performance regressions in practical deployments. 
The reconstruction quality of SAEs was tested using multiple metrics: (i) the cosine similarity between the input and the reconstructed output; (ii) the downstream performance (Recall@20 and nDCG@20) of the CFAE with a nested SAE, relative to the performance of the unmodified CFAE; and (iii) the average number of neurons firing per input (denoted $\mathrm{L_0}$).



Figure~\ref{fig2:sparsity_vs_accuracy} presents the results for SAEs based on 1024-dimensional ELSA and MultVAE backbones, evaluated on the test users of the ML-25M dataset.\footnote{Additional results are available from the online repository.}
Our experiments revealed that CFAEs with nested SAEs can retain high performance relative to their original counterparts, though robustness varies across SAE and CFAE architectures. In particular, TopK SAEs consistently achieved a more favorable sparsity–-accuracy trade-off compared to Basic SAEs. The $\lambda_1$ hyperparameter, which controls the sparsity of Basic SAE latent representations, is difficult to tune and highly sensitive; even small misconfigurations can significantly degrade either sparsity or reconstruction quality. In contrast, TopK SAEs explicitly constrain the number of active neurons, making them easier to configure and more effective overall.

For CFAEs that use cosine-like embeddings, TopK SAEs can reconstruct the latent space with minimal degradation in downstream performance. For example, on ML-25M, 1024-dimensional ELSA embeddings could be reconstructed using as few as $k=8$ neurons, while preserving 87\% of the original model’s nDCG@20. When the number of active neurons was increased to $k = 32$, the nested autoencoder recovered 95\% of the original Recall@20 and 94\% of nDCG@20, with a reconstruction cosine similarity of 90\%.

Our ablation study further showed that replacing the $\mathrm{L}_2$ reconstruction loss (see Eq. \eqref{sae_loss}) with a negative cosine similarity loss,

\begin{align*}
    1 - \Big(\frac{y_i}{\|y_i\|}\Big)^T\Big(\frac{\tilde{y}_i}{\|\tilde{y}_i\|}\Big), \quad \text{where} \quad \tilde{y}_i = (\mathcal{D}_s \circ \mathcal{E}_s)(y_i),
\end{align*}

\noindent improved the sparsity--accuracy trade-off for cosine-like embeddings, as used by ELSA. In the previously mentioned $k = 32$ case, this loss yielded $97\%$ of ELSA's Recall@20 and nDCG@20, with a cosine reconstruction similarity of $91\%$. However, in the case of MultVAE, this loss no longer ensures a small Euclidean distance between variational embeddings -- which represent samples from a probability distribution -- and their reconstructions. As a result, downstream performance degraded significantly.

Overall, our results suggest that CFAE models optimized for cosine similarity between embeddings are particularly well-suited to our two-stage autoencoder setup. Moreover, we hypothesize that the SAE approach may generalize to other non-autoencoder CF architectures (e.g., matrix factorization or graph neural networks) that rely on embedding angles -- rather than distances -- for inference. We encourage future work to explore this broader applicability.

\begin{table*}[tb] 
    \caption{Tags with highest (black) and lowest (grey) KL divergence from average activation distributions over all tags. \normalfont{All three models selected \textit{Quentin Tarantino} as one of the most diverse tags. Also \textit{James Bond} and \textit{Coen brothers} were listed by two out of three models.}}
    \centering
    \begin{subfigure}[t]{0.32\textwidth}
        \centering
        \begin{tabular}{lrr}
        \toprule
        Tag & $H$ & $D_{\mathrm{KL}}$ \\
        \midrule
james bond & 2.06 & 14.30 \\
quentin tarantino & 2.86 & 14.24 \\
studio ghibli & 1.67 & 14.16 \\
star trek & 1.64 & 14.01 \\
robert rodriguez & 1.90 & 14.00 \\[3pt]
\color{gray}boring & \color{gray}5.44 & \color{gray}5.76 \\
        \bottomrule
        \end{tabular}
        \caption{MultVAE + L2}
    \end{subfigure}
    \hfill 
\begin{subfigure}[t]{0.32\textwidth}
        \centering
        \begin{tabular}{lrr}
        \toprule
        Tag & $H$ & $D_{\mathrm{KL}}$ \\
        \midrule
quentin tarantino & 4.65 & 14.60 \\
wes anderson & 4.09 & 14.25 \\
stanley kubrick & 4.50 & 14.20 \\
monty python & 3.52 & 14.04 \\
coen brothers & 4.81 & 13.99 \\[3pt]
\color{gray}boring & \color{gray}7.33 & \color{gray}3.15 \\
        \bottomrule
        \end{tabular}        
        \caption{ELSA + L2}
    \end{subfigure}
    \hfill
\begin{subfigure}[t]{0.32\textwidth}
        \centering
        \begin{tabular}{lrr}
        \toprule
        Tag & $H$ & $D_{\mathrm{KL}}$ \\
        \midrule
quentin tarantino & 4.90 & 14.34 \\
coen brothers & 5.21 & 14.20 \\
james bond & 5.18 & 14.20 \\
tim burton & 5.28 & 14.10 \\
jim carrey & 5.17 & 14.07 \\[3pt]
\color{gray}bd-r & \color{gray}8.22 & \color{gray}2.65 \\
        \bottomrule
        \end{tabular}        
        \caption{ELSA + Cosine}
    \end{subfigure}

    \label{tab:tags}
\end{table*}

\begin{figure*}
    \centering
    \includegraphics[width=1.0\textwidth]{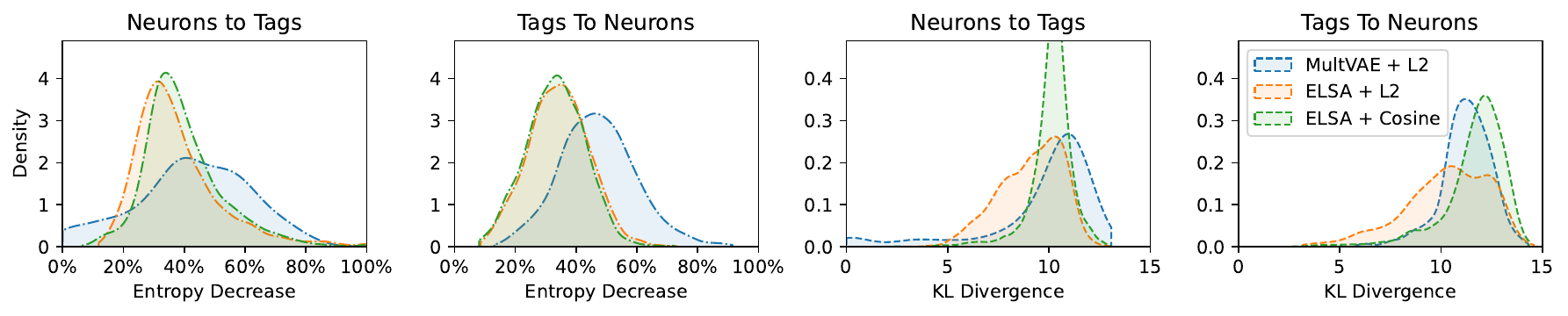}
    \caption{KL divergence and entropy decrease of tags and neurons. \normalfont{For ``tags to neurons'' direction, relative entropy decrease is defined as $(H-H_t)/H$, where $H_t$ is the entropy of tag's distribution of neuron activations, while $H$ is the entropy of average distribution over all tags. The other direction is defined analogically.}}
    \label{fig:rq2TagToNeuron}
\end{figure*}


\subsection{Interpretability of SAE Neurons}
\label{sec:intersae}

The next step toward SAE-based steering is to determine whether a correspondence has emerged between sparse neurons and semantic concepts -- that is, to identify which "knobs" can be manipulated to steer recommendations in a desired way. 

In this experiment, we utilized the TF-IDF based concept-neuron mapping as described in Section \ref{sec:concept_neuron_mapping} on top of ML-25M user-created textual tags. We preprocessed the dataset by removing tags that appear fewer than 100 times to reduce noise, and retained only those associated with items present in the interaction data. We analyzed three nested autoencoder variants based on CFAEs trained in our accuracy experiments. Two backbones were used: ELSA with a 1024-dimensional encoder, and MultVAE with the same encoder dimensionality. For both backbones, we trained a TopK SAE with an $\mathrm{L}_2$ reconstruction loss, an $8\times$ scaling factor (resulting in an 8,192-dimensional sparse layer), $k = 16$, and $\lambda_1 = 0.0003$. 
Additionally, we trained a second TopK SAE using the ELSA backbone with the same configuration but replacing the $\mathrm{L}_2$ loss with a cosine reconstruction loss. 
In the results, the three experimental configurations were denoted as \texttt{ELSA+L2}, \texttt{MultVAE+L2}, and \texttt{ELSA+Cosine}, respectively.


\begin{figure*}[tb]
    \centering    
    \includegraphics[width=0.85\textwidth]{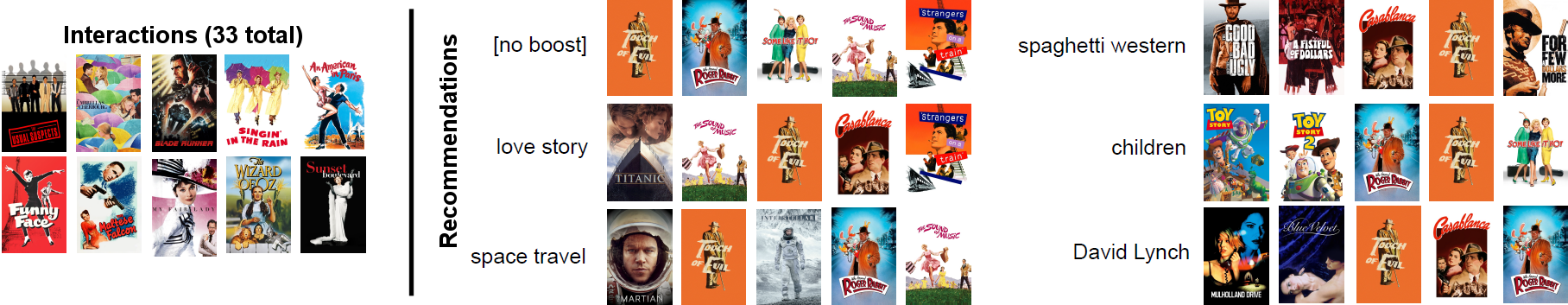}
    \caption{Effects of SAE-based steering on downstream recommendations for a particular user. 
    }
    \label{fig:steering}
\end{figure*}

To quantify the selectivity of neuron activation for individual tags, we analyzed the tag-to-neuron score matrix $M_{t \rightarrow n}$ using two distributional metrics. For each tag (i.e., row of $M_{t \rightarrow n}$, interpreted as a probability distribution via $\mathrm{L}_1$ normalization denoted by $\tilde{\cdot}$), we computed (1) the entropy $H$, which measures the spreading of activation across neurons, and (2) the Kullback-Leibler divergence $D_{\mathrm{KL}}$, which compares each tag $\tau_i$'s activation distribution $\tilde{S}_{t \to n}[\tau_i]$ to the average distribution across all tags $\mathbb{E}_\tau \left[M_{t \to n}[\tau]\right]$; formally $D_{\mathrm{KL}}\left( \mathbb{E}_\tau \left[\tilde{S}_{t \to n}[\tau] \right] \,\big\|\, \tilde{S}_{t \to n}[\tau_i] \right)$. Figure~\ref{fig:rq2TagToNeuron} presents the distributions of $D_{\mathrm{KL}}$ and relative entropy decrease across tags for all three model variants, while Table~\ref{tab:tags} lists tags with the most and least specific activation patterns. We found that a large portion of tags exhibit significantly lower entropy than the per-model averages of $H = 6.3$, $8.3$, and $9.0$ for \texttt{MultVAE+L2}, \texttt{ELSA+L2}, and \texttt{ELSA+Cosine}, respectively. Similarly, many tags showed highly divergent activation patterns (e.g., $88\%$ of \texttt{MultVAE+L2}, $59\%$ of \texttt{ELSA+L2}, and $88\%$ of \texttt{ELSA+Cosine} entries have $D_{\mathrm{KL}} > 10$). Tags with the highest $D_{\mathrm{KL}}$ typically correspond to individuals associated with movies (i.e., actors, characters, writers, or directors), while tags with the lowest divergence tend to be vague and less informative about the movies themselves, e.g., \emph{BD-R} (a type of optical media) or ``\emph{boring}''.

We applied an analogous analysis in the opposite direction -- from neurons to tags -- using $M_{n \rightarrow t}$. As shown in Figure~\ref{fig:rq2TagToNeuron}, many neurons exhibit large $D_{\mathrm{KL}}$ and relative entropy decrease, indicating sparse, targeted responses to a small subset of tags.

Comparing the models, \texttt{MultVAE+L2} achieved the highest relative entropy decrease across both directions, followed by \texttt{ELSA+Cosine} and then \texttt{ELSA+L2}. For $D_{\mathrm{KL}}$, \texttt{MultVAE+L2} and \texttt{ELSA+Cosine} showed similar selectivity, with \texttt{ELSA+L2} slightly trailing behind.

In summary, despite using only simple metadata for concept identification, we have shown that a significant portion of the learned neurons appears to be activated by compact, well-defined, and semantically meaningful features -- revealing a layer of control knobs with sufficient interpretability and fidelity.

\subsection{Practical Steering of Recommendations}
\label{sec:steeringEval}
We conclude our experimental study by evaluating whether the learned control panel of sparse neurons enables effective intervention in recommendation behavior. 
Let us start with a qualitative evaluation depicted in Figure~\ref{fig:steering}. Here, we employed the \texttt{ELSA+L2} nested autoencoder and $\text{argmax}_n M_{n \rightarrow t}$ concept-neuron mapping to steer the recommendations of a particular user towards several segments. In particular, the user originally interacted with 33 movies, and the default list of recommendations (denoted as \emph{[no boost]}) reflected their general interests centered on classic Hollywood films, musicals, and film noir. Boosting the characteristic neurons corresponding to selected concepts
(with $\alpha=0.15$) shifts the recommendations toward the desired themes while preserving the user's original preferences. For instance, amplifying neurons associated with the ``\emph{Children}'' and ``\emph{David Lynch}'' tags results in the inclusion of \emph{Toy Story} and \emph{Mulholland Drive}, respectively. Notably, steering behaves as a \textbf{blend} of the original profile and the targeted concept. For example, in addition to the popular \emph{Titanic}, amplifying the ``\emph{Love story}'' neuron also promotes \emph{The Sound of Music} and \emph{Casablanca}, reflecting a preference for romantic films within classic Hollywood and musical genres.

\begin{figure}[tb]
    \centering    
    \includegraphics[width=\linewidth]{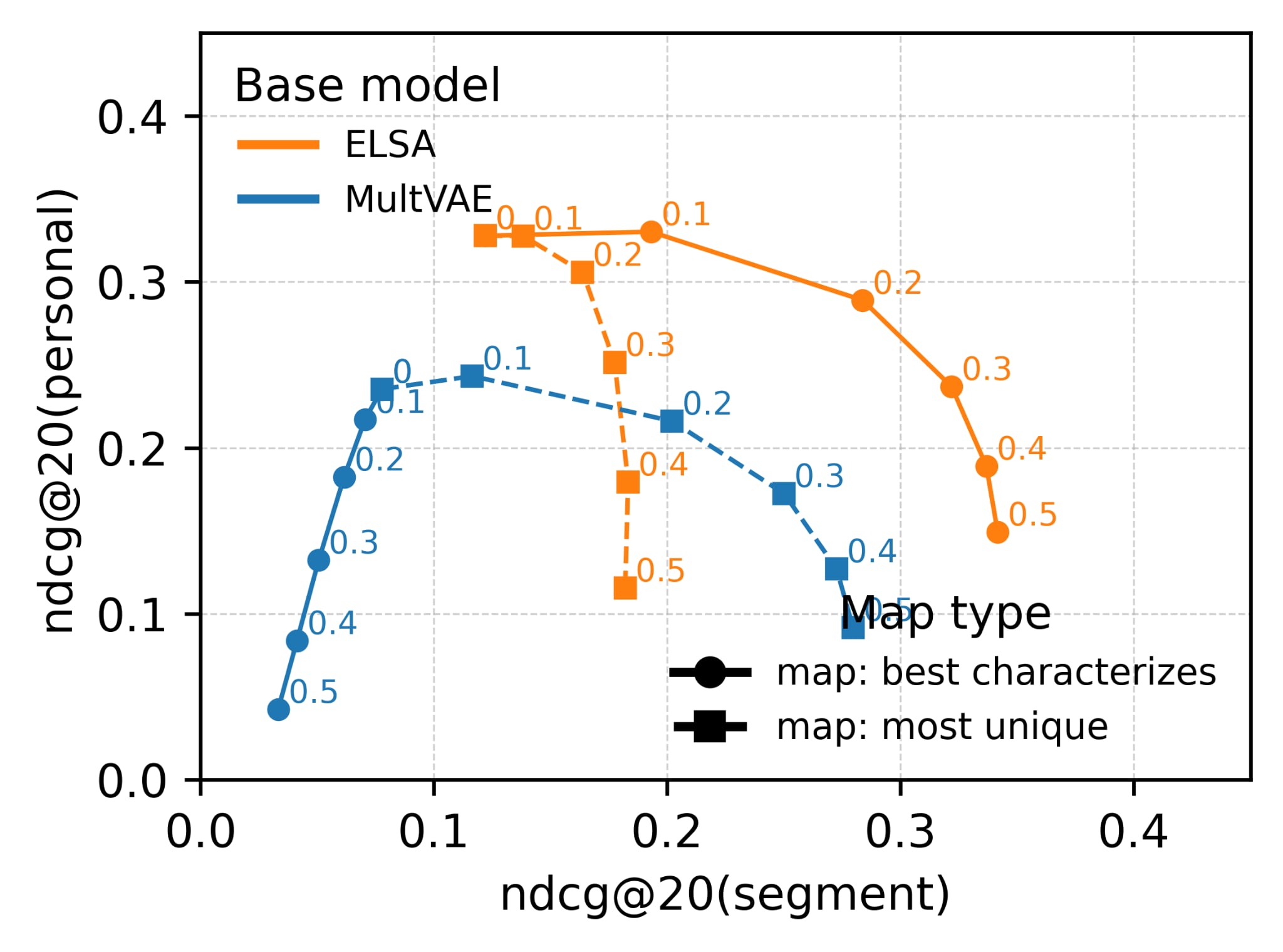}
    \caption{Results of the steering procedure for varying intensities $\alpha$ and two concept-neuron mappings.}
    \label{fig:steeringOverall}
\end{figure}

Let us now observe whether such a blend of original and segment-wise preferences can be obtained in general. For each user, we identified the segment $\mathcal{S}_u$ that stands out the most when comparing their hold-out sets against the inference input.\footnote{Here we define segments on top of user-assigned tags, i.e.,  item $i \in \text{segment } \mathcal{S}_t$ induced by a tag $t$ if $t$ has been assigned to $i$.} Then, we applied a steering procedure towards the segment $\mathcal{S}_u$ with gradually increasing $\alpha$. To map segments to neurons, we tested both $\text{argmax}_n M_{t \rightarrow n}$ (i.e., most \textit{unique} neuron) and $\text{argmax}_n M_{n \rightarrow t}$ (most \textit{representative} neuron) and observed the the tradeoff between the general recommendation relevance (w.r.t. user's hold-out set) and the ability to steer, i.e., recommend items from segment $\mathcal{S}_u$.

Figure \ref{fig:steeringOverall} depicts the results of \texttt{ELSA+L2} and \texttt{MultVAE+L2}. For \texttt{ELSA+L2} and mild steering ($\alpha=0.1$), both mapping strategies delivered relevant recommendations that are more aligned towards the target segment. Nevertheless, steering based on the most \textit{unique} neuron could only boost a fraction of the whole segment and delivers inferior relevance vs. steering tradeoff as compared to boosting the most \textit{representative} neuron. However, for \texttt{MultVAE+L2}, boosting \textit{representative} neurons completely failed to deliver steered recommendations due to large overlaps in $\text{argmax}_n M_{n \rightarrow t}$ mapping across tags.
Using \textit{unique} neurons partially remedied the issue, yet its results remained inferior to those of \texttt{ELSA+L2}.  

Overall, our experiments confirmed that steering based on SAE ``hooks'' is plausible. However, in contrast to previous work~\cite{wang2024interpretinternalstatesrecommendation}, we demonstrated that \emph{not all CF architectures} are equally suitable for such modifications, and some caution is warranted.   

\section{Conclusions and Future Work}\label{sec:conclusions}

In this work, we explored the applicability of mechanistic interpretability techniques to CF.
Specifically, we used SAEs inserted between the CFAE encoder and decoder to transform dense, entangled latent user representations into sparse, interpretable ones.
We evaluated (1) the robustness and downstream accuracy of sparse reconstruction, (2) interpretability of SAE embeddings, and (3) steerability of recommendations by boosting specific learned ``knobs''. Overall, embedding reconstruction was effective: TopK SAEs preserved almost all downstream accuracy for cosine-like CFAE embeddings, while Basic SAEs required careful hyperparameter tuning. Additionally, we found that simple structured metadata and processing can yield interpretable concept-to-neuron mappings.

As a foundational step toward SAE-based steering of CF recommendations, this work has several limitations.
We evaluated relatively simple SAE variants, leaving room for improved accuracy via wider sparse layers or auxiliary losses (e.g., \cite{gao2024scalingevaluatingsparseautoencoders}) and more advanced SAE architectures, e.g., \cite{bussmann2024batchtopksparseautoencoders,pmlr-v267-bussmann25a}. Moreover, we only experimented with coarse segments and concept-neuron mappings based on item metadata. Applying large language models to full-text item descriptions could reveal finer neuron–item relationships, and examining neuron activations on multi-item inputs may surface additional inter-item relational features useful for steering and explanation. Finally, we showed that not all embedding methods are robust enough for sparse reconstruction and steering, yet embeddings based on cosine similarity proved highly robust in our experiments. Thus, we believe our pipeline can be easily expanded to, e.g., matrix factorizations or graph neural networks, and plan to verify this hypothesis in future work.


\begin{acks}
This paper has been supported by the Czech Science Foundation (GA\v{C}R) project 25-16785S.
\end{acks}


\bibliographystyle{ACM-Reference-Format}
\bibliography{sample-base}

@InProceedings{pmlr-v267-bussmann25a,
  title = 	 {Learning Multi-Level Features with Matryoshka Sparse Autoencoders},
  author =       {Bussmann, Bart and Nabeshima, Noa and Karvonen, Adam and Nanda, Neel},
  booktitle = 	 {Proceedings of the 42nd International Conference on Machine Learning},
  pages = 	 {6077--6101},
  year = 	 {2025},
  editor = 	 {Singh, Aarti and Fazel, Maryam and Hsu, Daniel and Lacoste-Julien, Simon and Berkenkamp, Felix and Maharaj, Tegan and Wagstaff, Kiri and Zhu, Jerry},
  volume = 	 {267},
  series = 	 {Proceedings of Machine Learning Research},
  month = 	 {13--19 Jul},
  publisher =    {PMLR},
  url = 	 {https://proceedings.mlr.press/v267/bussmann25a.html}
}

@Inbook{Koren2022,
author="Koren, Yehuda
and Rendle, Steffen
and Bell, Robert",
editor="Ricci, Francesco
and Rokach, Lior
and Shapira, Bracha",
title="Advances in Collaborative Filtering",
bookTitle="Recommender Systems Handbook",
year="2022",
publisher="Springer US",
address="New York, NY",
pages="91--142",
isbn="978-1-0716-2197-4",
doi="10.1007/978-1-0716-2197-4_3",
url="https://doi.org/10.1007/978-1-0716-2197-4_3"
}

@inproceedings{10.1145/3640457.3688138,
author = {Ganh\"{o}r, Christian and Moscati, Marta and Hausberger, Anna and Nawaz, Shah and Schedl, Markus},
title = {A Multimodal Single-Branch Embedding Network for Recommendation in Cold-Start and Missing Modality Scenarios},
year = {2024},
isbn = {9798400705052},
publisher = {Association for Computing Machinery},
address = {New York, NY, USA},
url = {https://doi.org/10.1145/3640457.3688138},
doi = {10.1145/3640457.3688138},
booktitle = {Proceedings of the 18th ACM Conference on Recommender Systems},
pages = {380–390},
numpages = {11},
location = {Bari, Italy},
series = {RecSys '24}
}

@inproceedings{10.1145/2959100.2959190,
author = {Covington, Paul and Adams, Jay and Sargin, Emre},
title = {Deep Neural Networks for YouTube Recommendations},
year = {2016},
isbn = {9781450340359},
publisher = {Association for Computing Machinery},
address = {New York, NY, USA},
url = {https://doi.org/10.1145/2959100.2959190},
doi = {10.1145/2959100.2959190},
booktitle = {Proceedings of the 10th ACM Conference on Recommender Systems},
pages = {191–198},
numpages = {8},
location = {Boston, Massachusetts, USA},
series = {RecSys '16}
}

@article{10.1145/3568022,
author = {Gao, Chen and Zheng, Yu and Li, Nian and Li, Yinfeng and Qin, Yingrong and Piao, Jinghua and Quan, Yuhan and Chang, Jianxin and Jin, Depeng and He, Xiangnan and Li, Yong},
title = {A Survey of Graph Neural Networks for Recommender Systems: Challenges, Methods, and Directions},
year = {2023},
issue_date = {March 2023},
publisher = {Association for Computing Machinery},
address = {New York, NY, USA},
volume = {1},
number = {1},
url = {https://doi.org/10.1145/3568022},
doi = {10.1145/3568022},
journal = {ACM Trans. Recomm. Syst.},
month = mar,
articleno = {3},
numpages = {51}
}

@Inbook{Nikolakopoulos2022,
author="Nikolakopoulos, Athanasios N.
and Ning, Xia
and Desrosiers, Christian
and Karypis, George",
editor="Ricci, Francesco
and Rokach, Lior
and Shapira, Bracha",
title="Trust Your Neighbors: A Comprehensive Survey of Neighborhood-Based Methods for Recommender Systems",
bookTitle="Recommender Systems Handbook",
year="2022",
publisher="Springer US",
address="New York, NY",
pages="39--89",
isbn="978-1-0716-2197-4",
doi="10.1007/978-1-0716-2197-4_2",
url="https://doi.org/10.1007/978-1-0716-2197-4_2"
}

@article{mcinnes2018umap,
  title={UMAP: Uniform Manifold Approximation and Projection for Dimension Reduction},
  author={McInnes, Leland and Healy, John and Melville, James},
  journal={arXiv preprint arXiv:1802.03426},
  year={2018}
}

@inproceedings{10.1145/3477495.3531890,
author = {Gao, Zhaolin and Shen, Tianshu and Mai, Zheda and Bouadjenek, Mohamed Reda and Waller, Isaac and Anderson, Ashton and Bodkin, Ron and Sanner, Scott},
title = {Mitigating the Filter Bubble While Maintaining Relevance: Targeted Diversification with VAE-based Recommender Systems},
year = {2022},
isbn = {9781450387323},
publisher = {Association for Computing Machinery},
address = {New York, NY, USA},
url = {https://doi.org/10.1145/3477495.3531890},
doi = {10.1145/3477495.3531890},
booktitle = {Proceedings of the 45th International ACM SIGIR Conference on Research and Development in Information Retrieval},
pages = {2524–2531},
numpages = {8},
location = {Madrid, Spain},
series = {SIGIR '22}
}

@article{jeunen2022embarrassingly,
  title={Embarrassingly shallow auto-encoders for dynamic collaborative filtering},
  author={Jeunen, Olivier and Van Balen, Jan and Goethals, Bart},
  journal={User Modeling and User-Adapted Interaction},
  volume={32},
  number={4},
  pages={509--541},
  year={2022},
  publisher={Springer}
}

@inproceedings{10.1145/3340631.3394864,
author = {Lu, Feng and Dumitrache, Anca and Graus, David},
title = {Beyond Optimizing for Clicks: Incorporating Editorial Values in News Recommendation},
year = {2020},
isbn = {9781450368612},
publisher = {Association for Computing Machinery},
address = {New York, NY, USA},
url = {https://doi.org/10.1145/3340631.3394864},
doi = {10.1145/3340631.3394864},
booktitle = {Proceedings of the 28th ACM Conference on User Modeling, Adaptation and Personalization},
pages = {145–153},
numpages = {9},
location = {Genoa, Italy},
series = {UMAP '20}
}

@inproceedings{10.1145/3687151.3687152,
author = {Kruse, Johannes and Lindskow, Kasper and Kalloori, Saikishore and Polignano, Marco and Pomo, Claudio and Srivastava, Abhishek and Uppal, Anshuk and Andersen, Michael Riis and Frellsen, Jes},
title = {EB-NeRD a large-scale dataset for news recommendation},
year = {2024},
isbn = {9798400711275},
publisher = {Association for Computing Machinery},
address = {New York, NY, USA},
url = {https://doi.org/10.1145/3687151.3687152},
doi = {10.1145/3687151.3687152},
booktitle = {Proceedings of the Recommender Systems Challenge 2024},
pages = {1–11},
numpages = {11},
location = {Bari, Italy},
series = {RecSysChallenge '24}
}

@Article{Mauro2023,
author={Mauro, Noemi
and Hu, Zhongli Filippo
and Ardissono, Liliana},
title={Justification of recommender systems results: a service-based approach},
journal={User Modeling and User-Adapted Interaction},
year={2023},
month={Jul},
day={01},
volume={33},
number={3},
pages={643-685},
issn={1573-1391},
doi={10.1007/s11257-022-09345-8},
url={https://doi.org/10.1007/s11257-022-09345-8}
}

@ARTICLE{4648950,
  author={Symeonidis, Panagiotis and Nanopoulos, Alexandros and Manolopoulos, Yannis},
  journal={IEEE Transactions on Systems, Man, and Cybernetics - Part A: Systems and Humans}, 
  title={Providing Justifications in Recommender Systems}, 
  year={2008},
  volume={38},
  number={6},
  pages={1262-1272},
  doi={10.1109/TSMCA.2008.2003969}}

@inproceedings{muhammad2016use,
  title={On the use of opinionated explanations to rank and justify recommendations},
  author={Muhammad, Khalil and Lawlor, Aonghus and Smyth, Barry},
  booktitle={The Twenty-Ninth International Flairs Conference},
  year={2016}
}

@inproceedings{Balog_2023, series={CHI ’23},
   title={Measuring the Impact of Explanation Bias: A Study of Natural Language Justifications for Recommender Systems},
   url={http://dx.doi.org/10.1145/3544549.3585748},
   DOI={10.1145/3544549.3585748},
   booktitle={Extended Abstracts of the 2023 CHI Conference on Human Factors in Computing Systems},
   publisher={ACM},
   author={Balog, Krisztian and Radlinski, Filip and Petrov, Andrey},
   year={2023},
   month=apr, pages={1–8},
   collection={CHI ’23} }

@misc{park2020jrecsprincipledscalablerecommendation,
      title={J-Recs: Principled and Scalable Recommendation Justification}, 
      author={Namyong Park and Andrey Kan and Christos Faloutsos and Xin Luna Dong},
      year={2020},
      eprint={2011.05928},
      archivePrefix={arXiv},
      primaryClass={cs.IR},
      url={https://arxiv.org/abs/2011.05928}, 
}

@misc{guesmi2023justificationvstransparencyvisual,
      title={Justification vs. Transparency: Why and How Visual Explanations in a Scientific Literature Recommender System}, 
      author={Mouadh Guesmi and Mohamed Amine Chatti and Shoeb Joarder and Qurat Ul Ain and Clara Siepmann and Hoda Ghanbarzadeh and Rawaa Alatrash},
      year={2023},
      eprint={2305.17034},
      archivePrefix={arXiv},
      primaryClass={cs.IR},
      url={https://arxiv.org/abs/2305.17034}, 
}

@misc{wang2024interpretinternalstatesrecommendation,
      title={Interpret the Internal States of Recommendation Model with Sparse Autoencoder}, 
      author={Jiayin Wang and Xiaoyu Zhang and Weizhi Ma and Min Zhang},
      year={2024},
      eprint={2411.06112},
      archivePrefix={arXiv},
      primaryClass={cs.IR},
      url={https://arxiv.org/abs/2411.06112}, 
}

@INPROCEEDINGS{8594844,
  author={Kang, Wang-Cheng and McAuley, Julian},
  booktitle={2018 IEEE International Conference on Data Mining (ICDM)}, 
  title={Self-Attentive Sequential Recommendation}, 
  year={2018},
  volume={},
  number={},
  pages={197-206},
  doi={10.1109/ICDM.2018.00035}}

@inproceedings{10.5555/3495724.3497368,
author = {Steck, Harald},
title = {Autoencoders that don't overfit towards the identity},
year = {2020},
isbn = {9781713829546},
publisher = {Curran Associates Inc.},
address = {Red Hook, NY, USA},
booktitle = {Proceedings of the 34th International Conference on Neural Information Processing Systems},
articleno = {1644},
numpages = {11},
location = {Vancouver, BC, Canada},
series = {NIPS '20}
}

@misc{depauw2022modellingusersitemmetadata,
      title={Modelling Users with Item Metadata for Explainable and Interactive Recommendation}, 
      author={Joey De Pauw and Koen Ruymbeek and Bart Goethals},
      year={2022},
      eprint={2207.00350},
      archivePrefix={arXiv},
      primaryClass={cs.IR},
      url={https://arxiv.org/abs/2207.00350}, 
}

@article{10.1145/3663364,
author = {Liang, Shangsong and Pan, Zhou and liu, wei and Yin, Jian and de Rijke, Maarten},
title = {A Survey on Variational Autoencoders in Recommender Systems},
year = {2024},
issue_date = {October 2024},
publisher = {Association for Computing Machinery},
address = {New York, NY, USA},
volume = {56},
number = {10},
issn = {0360-0300},
url = {https://doi.org/10.1145/3663364},
doi = {10.1145/3663364},
journal = {ACM Comput. Surv.},
month = jun,
articleno = {268},
numpages = {40}
}

@inproceedings{10.1145/3298689.3347015,
author = {Kim, Daeryong and Suh, Bongwon},
title = {Enhancing VAEs for collaborative filtering: flexible priors \& gating mechanisms},
year = {2019},
isbn = {9781450362436},
publisher = {Association for Computing Machinery},
address = {New York, NY, USA},
url = {https://doi.org/10.1145/3298689.3347015},
doi = {10.1145/3298689.3347015},
booktitle = {Proceedings of the 13th ACM Conference on Recommender Systems},
pages = {403–407},
numpages = {5},
location = {Copenhagen, Denmark},
series = {RecSys '19}
}

@inproceedings{absease,
author = {Spi\v{s}\'{a}k, Martin and Bartyzal, Radek and Hoskovec, Anton\'{\i}n and Pe\v{s}ka, Ladislav},
title = {On Interpretability of Linear Autoencoders},
year = {2024},
isbn = {9798400705052},
publisher = {Association for Computing Machinery},
address = {New York, NY, USA},
url = {https://doi.org/10.1145/3640457.3688179},
doi = {10.1145/3640457.3688179},
booktitle = {Proceedings of the 18th ACM Conference on Recommender Systems},
pages = {975–980},
numpages = {6},
location = {Bari, Italy},
series = {RecSys '24}
}

@inproceedings{BertinMahieux2011TheMS,
  author       = {Thierry Bertin{-}Mahieux and
                  Daniel P. W. Ellis and
                  Brian Whitman and
                  Paul Lamere},
  editor       = {Anssi Klapuri and
                  Colby Leider},
  title        = {The Million Song Dataset},
  booktitle    = {Proceedings of the 12th International Society for Music Information
                  Retrieval Conference, {ISMIR} 2011, Miami, Florida, USA, October 24-28,
                  2011},
  pages        = {591--596},
  publisher    = {University of Miami},
  address = {Miami, FL, USA},
  year         = {2011},
  url          = {http://ismir2011.ismir.net/papers/OS6-1.pdf},
}

@inproceedings{sedhain2015autorec,
  title={Autorec: Autoencoders meet collaborative filtering},
  author={Sedhain, Suvash and Menon, Aditya Krishna and Sanner, Scott and Xie, Lexing},
  booktitle={Proceedings of the 24th international conference on World Wide Web},
  pages={111--112},
  year={2015}
}

@inproceedings{vanvcura2024beeformer,
  title={beeFormer: Bridging the Gap Between Semantic and Interaction Similarity in Recommender Systems},
  author={Van{\v{c}}ura, Vojt{\v{e}}ch and Kord{\'\i}k, Pavel and Straka, Milan},
  booktitle={Proceedings of the 18th ACM Conference on Recommender Systems},
  pages={1102--1107},
  year={2024}
}

@misc{bereska2024mechanisticinterpretabilityaisafety,
      title={Mechanistic Interpretability for AI Safety -- A Review}, 
      author={Leonard Bereska and Efstratios Gavves},
      year={2024},
      eprint={2404.14082},
      archivePrefix={arXiv},
      primaryClass={cs.AI},
      url={https://arxiv.org/abs/2404.14082}, 
}

@misc{cunningham2023sparseautoencodershighlyinterpretable,
      title={Sparse Autoencoders Find Highly Interpretable Features in Language Models}, 
      author={Hoagy Cunningham and Aidan Ewart and Logan Riggs and Robert Huben and Lee Sharkey},
      year={2023},
      eprint={2309.08600},
      archivePrefix={arXiv},
      primaryClass={cs.LG},
      url={https://arxiv.org/abs/2309.08600}, 
}

@article{bricken2023towards,
  title={Towards monosemanticity: Decomposing language models with dictionary learning},
  author={Bricken, Trenton and Templeton, Adly and Batson, Joshua and Chen, Brian and Jermyn, Adam and Conerly, Tom and Turner, Nick and Anil, Cem and Denison, Carson and Askell, Amanda and others},
  journal={Transformer Circuits Thread},
  volume={2},
  year={2023}
}

@misc{templeton2024scaling,
  title={Scaling monosemanticity: Extracting interpretable features from claude 3 sonnet. Transformer Circuits Thread},
  author={Templeton, Adly and Conerly, Tom and Marcus, Jonathan and Lindsey, Jack and Bricken, Trenton and Chen, Brian and Pearce, Adam and Citro, Craig and Ameisen, Emmanuel and Jones, Andy and others},
  year={2024}
}

@misc{makhzani2014ksparseautoencoders,
      title={k-Sparse Autoencoders}, 
      author={Alireza Makhzani and Brendan Frey},
      year={2014},
      eprint={1312.5663},
      archivePrefix={arXiv},
      primaryClass={cs.LG},
      url={https://arxiv.org/abs/1312.5663}, 
}

@misc{gao2024scalingevaluatingsparseautoencoders,
      title={Scaling and evaluating sparse autoencoders}, 
      author={Leo Gao and Tom Dupré la Tour and Henk Tillman and Gabriel Goh and Rajan Troll and Alec Radford and Ilya Sutskever and Jan Leike and Jeffrey Wu},
      year={2024},
      eprint={2406.04093},
      archivePrefix={arXiv},
      primaryClass={cs.LG},
      url={https://arxiv.org/abs/2406.04093}, 
}

@inproceedings{guo2024dualvae,
  title={DualVAE: Dual Disentangled Variational AutoEncoder for Recommendation},
  author={Guo, Zhiqiang and Li, Guohui and Li, Jianjun and Wang, Chaoyang and Shi, Si},
  booktitle={Proceedings of the 2024 SIAM International Conference on Data Mining (SDM)},
  pages={571--579},
  year={2024},
  organization={SIAM}
}

@inproceedings{refinetti2022dynamics,
  title={The dynamics of representation learning in shallow, non-linear autoencoders},
  author={Refinetti, Maria and Goldt, Sebastian},
  booktitle={International Conference on Machine Learning},
  pages={18499--18519},
  year={2022},
  organization={PMLR}
}

@inproceedings{vsafavrik2022repsys,
  title={Repsys: Framework for interactive evaluation of recommender systems},
  author={{\v{S}}afa{\v{r}}{\'\i}k, Jan and Van{\v{c}}ura, Vojt{\v{e}}ch and Kord{\'\i}k, Pavel},
  booktitle={Proceedings of the 16th ACM Conference on Recommender Systems},
  pages={636--639},
  year={2022}
}

@inproceedings{10.1145/3523227.3551482,
author = {Van\v{c}ura, Vojt\v{e}ch and Alves, Rodrigo and Kasalick\'{y}, Petr and Kord\'{\i}k, Pavel},
title = {Scalable Linear Shallow Autoencoder for Collaborative Filtering},
year = {2022},
isbn = {9781450392785},
publisher = {Association for Computing Machinery},
address = {New York, NY, USA},
url = {https://doi.org/10.1145/3523227.3551482},
doi = {10.1145/3523227.3551482},
booktitle = {Proceedings of the 16th ACM Conference on Recommender Systems},
pages = {604–609},
numpages = {6},
location = {Seattle, WA, USA},
series = {RecSys '22}
}

@article{10.1145/2827872,
author = {Harper, F. Maxwell and Konstan, Joseph A.},
title = {The MovieLens Datasets: History and Context},
year = {2015},
issue_date = {January 2016},
publisher = {Association for Computing Machinery},
address = {New York, NY, USA},
volume = {5},
number = {4},
issn = {2160-6455},
url = {https://doi.org/10.1145/2827872},
doi = {10.1145/2827872},
journal = {ACM Trans. Interact. Intell. Syst.},
month = {dec},
articleno = {19},
numpages = {19}
}

@misc{kingma2017adammethodstochasticoptimization,
      title={Adam: A Method for Stochastic Optimization}, 
      author={Diederik P. Kingma and Jimmy Ba},
      year={2017},
      eprint={1412.6980},
      archivePrefix={arXiv},
      primaryClass={cs.LG},
      url={https://arxiv.org/abs/1412.6980}, 
}

@inproceedings{10.1145/3178876.3186150,
author = {Liang, Dawen and Krishnan, Rahul G. and Hoffman, Matthew D. and Jebara, Tony},
title = {Variational Autoencoders for Collaborative Filtering},
year = {2018},
isbn = {9781450356398},
publisher = {International World Wide Web Conferences Steering Committee},
address = {Republic and Canton of Geneva, CHE},
url = {https://doi.org/10.1145/3178876.3186150},
doi = {10.1145/3178876.3186150},
booktitle = {Proceedings of the 2018 World Wide Web Conference},
pages = {689–698},
numpages = {10},
location = {Lyon, France},
series = {WWW '18}
}

@inproceedings{10.1145/3336191.3371831,
author = {Shenbin, Ilya and Alekseev, Anton and Tutubalina, Elena and Malykh, Valentin and Nikolenko, Sergey I.},
title = {RecVAE: A New Variational Autoencoder for Top-N Recommendations with Implicit Feedback},
year = {2020},
isbn = {9781450368223},
publisher = {Association for Computing Machinery},
address = {New York, NY, USA},
url = {https://doi.org/10.1145/3336191.3371831},
doi = {10.1145/3336191.3371831},
booktitle = {Proceedings of the 13th International Conference on Web Search and Data Mining},
pages = {528–536},
numpages = {9},
location = {Houston, TX, USA},
series = {WSDM '20}
}

@inproceedings{10.1109/ICDM.2011.134,
author = {Ning, Xia and Karypis, George},
title = {SLIM: Sparse Linear Methods for Top-N Recommender Systems},
year = {2011},
isbn = {9780769544083},
publisher = {IEEE Computer Society},
address = {USA},
url = {https://doi.org/10.1109/ICDM.2011.134},
doi = {10.1109/ICDM.2011.134},
booktitle = {Proceedings of the 2011 IEEE 11th International Conference on Data Mining},
pages = {497–506},
numpages = {10},
series = {ICDM '11}
}

@inproceedings{sansa,
author = {Spi\v{s}\'{a}k, Martin and Bartyzal, Radek and Hoskovec, Anton\'{\i}n and Peska, Ladislav and T\r{u}ma, Miroslav},
title = {Scalable Approximate NonSymmetric Autoencoder for Collaborative Filtering},
year = {2023},
isbn = {9798400702419},
publisher = {Association for Computing Machinery},
address = {New York, NY, USA},
url = {https://doi.org/10.1145/3604915.3608827},
doi = {10.1145/3604915.3608827},
booktitle = {Proceedings of the 17th ACM Conference on Recommender Systems},
pages = {763–770},
numpages = {8},
location = {Singapore, Singapore},
series = {RecSys '23}
}

@inproceedings{10.5555/3454287.3454778,
author = {Steck, Harald},
title = {Markov Random Fields for Collaborative Filtering},
year = {2019},
publisher = {Curran Associates Inc.},
address = {Red Hook, NY, USA},
booktitle = {Proceedings of the 33rd International Conference on Neural Information Processing Systems},
articleno = {491},
numpages = {12}
}

@inproceedings{10.1145/3523227.3546772,
author = {Liang, Yu and Willemsen, Martijn C.},
title = {Exploring the longitudinal effects of nudging on users’ music genre exploration behavior and listening preferences},
year = {2022},
isbn = {9781450392785},
publisher = {Association for Computing Machinery},
address = {New York, NY, USA},
url = {https://doi.org/10.1145/3523227.3546772},
doi = {10.1145/3523227.3546772},
booktitle = {Proceedings of the 16th ACM Conference on Recommender Systems},
pages = {3–13},
numpages = {11},
location = {Seattle, WA, USA},
series = {RecSys '22}
}

@misc{bussmann2024batchtopksparseautoencoders,
      title={BatchTopK Sparse Autoencoders}, 
      author={Bart Bussmann and Patrick Leask and Neel Nanda},
      year={2024},
      eprint={2412.06410},
      archivePrefix={arXiv},
      primaryClass={cs.LG},
      url={https://arxiv.org/abs/2412.06410}, 
}

@article{liu2018novel,
  title={A novel deep hybrid recommender system based on auto-encoder with neural collaborative filtering},
  author={Liu, Yu and Wang, Shuai and Khan, M Shahrukh and He, Jieyu},
  journal={Big Data Mining and Analytics},
  volume={1},
  number={3},
  pages={211--221},
  year={2018},
  publisher={TUP}
}

@inproceedings{li2015deep,
  title={Deep collaborative filtering via marginalized denoising auto-encoder},
  author={Li, Sheng and Kawale, Jaya and Fu, Yun},
  booktitle={Proceedings of the 24th ACM international on conference on information and knowledge management},
  pages={811--820},
  year={2015}
}

@article{li2024survey,
  title={A survey on deep neural networks in collaborative filtering recommendation systems},
  author={Li, Pang and Noah, Shahrul Azman Mohd and Sarim, Hafiz Mohd},
  journal={arXiv preprint arXiv:2412.01378},
  year={2024}
}

@inproceedings{10.1145/3308558.3313710,
author = {Steck, Harald},
title = {Embarrassingly Shallow Autoencoders for Sparse Data},
year = {2019},
isbn = {9781450366748},
publisher = {Association for Computing Machinery},
address = {New York, NY, USA},
url = {https://doi.org/10.1145/3308558.3313710},
doi = {10.1145/3308558.3313710},
booktitle = {The World Wide Web Conference},
pages = {3251–3257},
numpages = {7},
location = {San Francisco, CA, USA},
series = {WWW '19}
}

@article{vanvcura2025evaluating,
  title={Evaluating Linear Shallow Autoencoders on Large Scale Datasets},
  author={Van{\v{c}}ura, Vojt{\v{e}}ch and Kasalick{\`y}, Petr and Alves, Rodrigo and Kord{\'\i}k, Pavel},
  journal={ACM Transactions on Recommender Systems},
  year={2025},
  publisher={ACM New York, NY}
}

@inproceedings{vanvcura2023scalable,
  title={Scalable and Explainable Linear Shallow Autoencoders for Collaborative Filtering from Industrial Perspective},
  author={Van{\v{c}}ura, Vojt{\v{e}}ch},
  booktitle={Proceedings of the 31st ACM Conference on User Modeling, Adaptation and Personalization},
  pages={290--295},
  year={2023}
}

@article{abusitta2024survey,
  title={Survey on Explainable AI: Techniques, challenges and open issues},
  author={Abusitta, Adel and Li, Miles Q and Fung, Benjamin CM},
  journal={Expert Systems with Applications},
  volume={255},
  pages={124710},
  year={2024},
  publisher={Elsevier}
}

@inproceedings{longo2020explainable,
  title={Explainable artificial intelligence: Concepts, applications, research challenges and visions},
  author={Longo, Luca and Goebel, Randy and Lecue, Freddy and Kieseberg, Peter and Holzinger, Andreas},
  booktitle={International cross-domain conference for machine learning and knowledge extraction},
  pages={1--16},
  year={2020},
  organization={Springer}
}

@inproceedings{xu2019explainable,
  title={Explainable AI: A brief survey on history, research areas, approaches and challenges},
  author={Xu, Feiyu and Uszkoreit, Hans and Du, Yangzhou and Fan, Wei and Zhao, Dongyan and Zhu, Jun},
  booktitle={Natural language processing and Chinese computing: 8th cCF international conference, NLPCC 2019, dunhuang, China, October 9--14, 2019, proceedings, part II 8},
  pages={563--574},
  year={2019},
  organization={Springer}
}

@inproceedings{10.1145/2792838.2800179,
author = {Harper, F. Maxwell and Xu, Funing and Kaur, Harmanpreet and Condiff, Kyle and Chang, Shuo and Terveen, Loren},
title = {Putting Users in Control of their Recommendations},
year = {2015},
isbn = {9781450336925},
publisher = {Association for Computing Machinery},
address = {New York, NY, USA},
url = {https://doi.org/10.1145/2792838.2800179},
doi = {10.1145/2792838.2800179},
booktitle = {Proceedings of the 9th ACM Conference on Recommender Systems},
pages = {3–10},
numpages = {8},
location = {Vienna, Austria},
series = {RecSys '15}
}

@inproceedings{10.1145/1357054.1357222,
author = {O'Donovan, John and Smyth, Barry and Gretarsson, Brynjar and Bostandjiev, Svetlin and H\"{o}llerer, Tobias},
title = {PeerChooser: visual interactive recommendation},
year = {2008},
isbn = {9781605580111},
publisher = {Association for Computing Machinery},
address = {New York, NY, USA},
url = {https://doi.org/10.1145/1357054.1357222},
doi = {10.1145/1357054.1357222},
booktitle = {Proceedings of the SIGCHI Conference on Human Factors in Computing Systems},
pages = {1085–1088},
numpages = {4},
location = {Florence, Italy},
series = {CHI '08}
}

@article{10.1080/10447318.2023.2262796,
author = {Sun, Ruixuan and Akella, Avinash and  Kong, Ruoyan and Zhou, Moyan and Konstan, Joseph A.},
title = {Interactive Content Diversity and User Exploration in Online Movie Recommenders: A Field Experiment},
journal = {International Journal of Human–Computer Interaction},
volume = {0},
number = {0},
pages = {1--15},
year = {2023},
publisher = {Taylor \& Francis},
doi = {10.1080/10447318.2023.2262796},


URL = { 
    
        https://doi.org/10.1080/10447318.2023.2262796
    
    

},
eprint = { 
    
        https://doi.org/10.1080/10447318.2023.2262796
    
    

}
}

@article{PARRA201543,
title = {User-controllable personalization: A case study with SetFusion},
journal = {International Journal of Human-Computer Studies},
volume = {78},
pages = {43-67},
year = {2015},
issn = {1071-5819},
doi = {https://doi.org/10.1016/j.ijhcs.2015.01.007},
url = {https://www.sciencedirect.com/science/article/pii/S1071581915000208},
author = {Denis Parra and Peter Brusilovsky}
}

@inproceedings{10.1145/3209219.3209223,
author = {Millecamp, Martijn and Htun, Nyi Nyi and Jin, Yucheng and Verbert, Katrien},
title = {Controlling Spotify Recommendations: Effects of Personal Characteristics on Music Recommender User Interfaces},
year = {2018},
isbn = {9781450355896},
publisher = {Association for Computing Machinery},
address = {New York, NY, USA},
url = {https://doi.org/10.1145/3209219.3209223},
doi = {10.1145/3209219.3209223},
booktitle = {Proceedings of the 26th Conference on User Modeling, Adaptation and Personalization},
pages = {101–109},
numpages = {9},
location = {Singapore, Singapore},
series = {UMAP '18}
}

@inproceedings{10.1145/2365952.2365964,
author = {Bostandjiev, Svetlin and O'Donovan, John and H\"{o}llerer, Tobias},
title = {TasteWeights: a visual interactive hybrid recommender system},
year = {2012},
isbn = {9781450312707},
publisher = {Association for Computing Machinery},
address = {New York, NY, USA},
url = {https://doi.org/10.1145/2365952.2365964},
doi = {10.1145/2365952.2365964},
booktitle = {Proceedings of the Sixth ACM Conference on Recommender Systems},
pages = {35–42},
numpages = {8},
location = {Dublin, Ireland},
series = {RecSys '12}
}

@inproceedings{10.1145/2365952.2365966,
author = {Knijnenburg, Bart P. and Bostandjiev, Svetlin and O'Donovan, John and Kobsa, Alfred},
title = {Inspectability and control in social recommenders},
year = {2012},
isbn = {9781450312707},
publisher = {Association for Computing Machinery},
address = {New York, NY, USA},
url = {https://doi.org/10.1145/2365952.2365966},
doi = {10.1145/2365952.2365966},
booktitle = {Proceedings of the Sixth ACM Conference on Recommender Systems},
pages = {43–50},
numpages = {8},
location = {Dublin, Ireland},
series = {RecSys '12}
}

@article{epstein2019generalization,
  title={Generalization bounds for unsupervised and semi-supervised learning with autoencoders},
  author={Epstein, Baruch and Meir, Ron},
  journal={arXiv preprint arXiv:1902.01449},
  year={2019}
}

@article{10.1093/idpl/ipx022,
    author = {Selbst, Andrew D and Powles, Julia},
    title = {Meaningful information and the right to explanation},
    journal = {International Data Privacy Law},
    volume = {7},
    number = {4},
    pages = {233-242},
    year = {2017},
    month = {12},
    issn = {2044-3994},
    doi = {10.1093/idpl/ipx022},
    url = {https://doi.org/10.1093/idpl/ipx022},
    eprint = {https://academic.oup.com/idpl/article-pdf/7/4/233/22923065/ipx022.pdf},
}

\appendix
\section{Reproducibility Details}
\label{sec:reproducibilityDetails}
We trained all CFAE variants using the Adam optimizer~\cite{kingma2017adammethodstochasticoptimization} with $\alpha = 3 \cdot 10^{-4}$ for ELSA and $10^{-3}$ for MultVAE, $\beta_1 = 0.9$, and $\beta_2 = 0.99$. For MultVAE, we annealed the $\beta$ parameter by $10^{-6}$ after each step. We trained both ELSA and MultVAE using a batch size of 1024 for up to 25 epochs. We employed early stopping for ELSA after 10 epochs without improvement in validation loss.

For the evaluation of SAE's reconstruction capability and downstream effectivity (Section \ref{sec:3_2}), we evaluated the following variants of Basic SAE and Top-k SAE: for both SAE architectures, we vary the width-to-input dimension ratios $\{2, 4, 8\}$ (corresponding to SAE embedding dimensions ranging from 1024 to 4096 for a 512-dimensional CFAE).
Apart from the embedding width, Basic SAE has a single hyperparameter -- the sparsity-inducing $\mathrm{L}_1$ coefficient -- which we vary over $\{0.0003, 0.001, 0.003, 0.01\}$.
For TopK SAE, we tune its sparsity-inducing hyperparameter $k$ for values in $\{8, 16, 32, 64\}$ and use a small $\mathrm{L}_1$ penalty of 0.0003 to prevent large spikes in activation values. 
ALL SAE variants were optimized using Adam~\cite{kingma2017adammethodstochasticoptimization} with $\alpha = 3 \cdot 10^{-4}, \beta_1 = 0.9$, and $\beta_2 = 0.99$.Training proceeded with a batch size of 1024 for up to 250 epochs, with early stopping triggered after 50 epochs without improvement in validation loss.

For the evaluation of concept-neuron mapping quality and steering experiments (Sections \ref{sec:intersae} and \ref{sec:steeringEval}), we utilized a more fine-grained training of selected SAE models with Adam ($\alpha = 1 \times 10^{-4}, \beta_1 = 0.9, \beta_2 = 0.99$), batch size of 1024 and up to 1,000 epochs with early stopping after 250 epochs without a validation loss improvement.

\section{Additional Experimental Results}
\label{sec:additionalExperiments}
\subsection{Detailed results of CFAE variants}
Table \ref{tab:cfae_results} depicts detailed results of individual CFAE variants w.r.t. Recall@20 and nDCG@20 metrics.
\begin{table}[h]
\centering
\caption{Accuracy of ELSA and MultVAE on ML-25M and MSD datasets. $d$ denotes the dimension of the bottleneck layer.}
\label{tab:cfae_results}
\begin{tabular}{lllrr}
\toprule
Dataset & Model    & $d$   & Recall@20                & nDCG@20                \\ 
\midrule
\multirow{6}{*}{ML-25M} 
        & ELSA     & 512   & $0.390 \pm 0.002$        & $0.353 \pm 0.002$      \\
        & ELSA     & 1024  & $0.396 \pm 0.002$        & $0.360 \pm 0.002$      \\
        & ELSA     & 2048  & $0.397 \pm 0.002$        & $0.363 \pm 0.002$      \\
\cmidrule(lr){2-5}
        & MultVAE  & 256   & $0.375 \pm 0.002$        & $0.333 \pm 0.002$      \\
        & MultVAE  & 512   & $0.383 \pm 0.002$        & $0.340 \pm 0.002$      \\
        & MultVAE  & 1024  & $0.377 \pm 0.002$        & $0.334 \pm 0.002$      \\
\midrule
\multirow{6}{*}{MSD} 
        & ELSA     & 512   & $0.245 \pm 0.001$        & $0.239 \pm 0.001$      \\
        & ELSA     & 1024  & $0.275 \pm 0.001$        & $0.269 \pm 0.001$      \\
        & ELSA     & 2048  & $0.298 \pm 0.001$        & $0.293 \pm 0.001$      \\
\cmidrule(lr){2-5}
        & MultVAE  & 256   & $0.211 \pm 0.001$        & $0.203 \pm 0.001$      \\
        & MultVAE  & 512   & $0.241 \pm 0.001$        & $0.232 \pm 0.001$      \\
        & MultVAE  & 1024  & $0.252 \pm 0.001$        & $0.242 \pm 0.001$      \\
\bottomrule
\end{tabular}
\end{table}

\begin{figure*}[h]
    \centering
    \includegraphics[width=\textwidth]{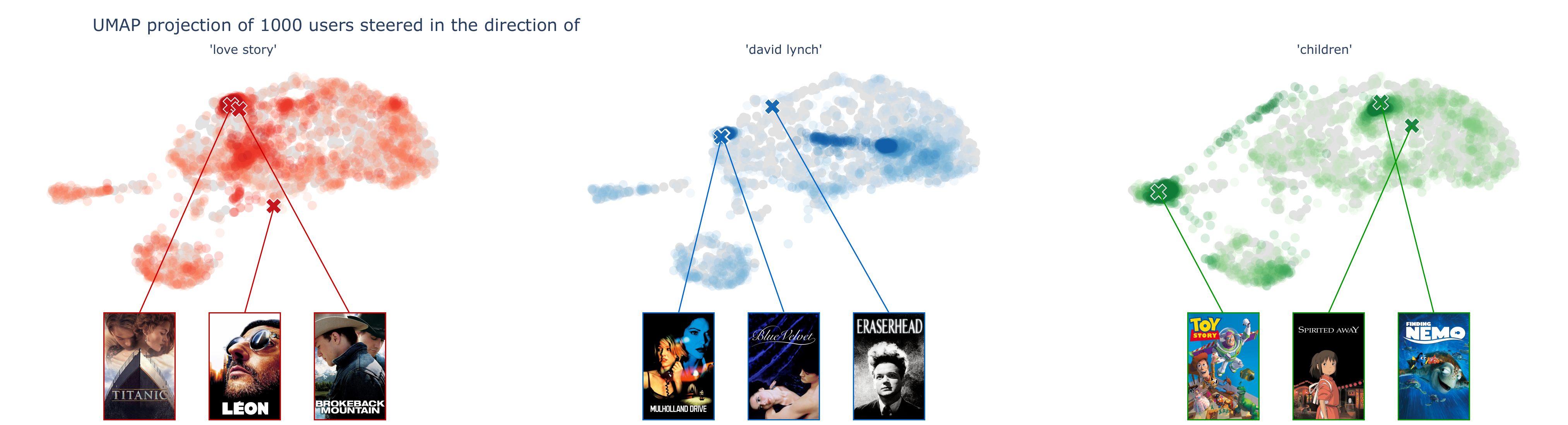}
    \caption{Effects of SAE-based steering on user representations. \normalfont{As adjustment strength increases (indicated by saturation), user embeddings shift toward regions associated with representative items of the boosted concept.}}
    \label{fig:steering_global}
\end{figure*}

\subsection{Distinctiveness of the Concept-Neuron Mappings}
In addition to the quantitative evaluation of the concept-neuron mapping, we can focus on the distinctiveness of such mappings. For this evaluation, we utilize the $\text{argmax}_t M_{n \rightarrow t}$ mapping, i.e., the most distinctive tags for each neuron, and, conversely, the most distinctive neurons for each tag ($\text{argmax}_n M_{t \rightarrow n}$). We observe that a substantial fraction of both tags and active neurons are identified as the most distinctive counterpart for at least one match. For example, in the \texttt{ELSA+L2} model, 834 unique neurons are identified as the most specific for at least one tag, and 985 unique tags are the most distinctive for at least one neuron. Overlaps -- where a neuron or tag is considered most distinctive for multiple counterparts -- often reflect meaningful relationships, such as correlated item-tag associations or duplicate metadata.
For instance, neuron \emph{id8156} is the most specific neuron for the tags \emph{superhero}, \emph{super-hero}, and \emph{comic book}; neuron  \emph{id1042} for \emph{archaeology} and \emph{indiana jones}, and neuron \emph{id2491} for \emph{ralph fiennes}, \emph{anthony hopkins}, \emph{edward norton}, \emph{fbi}, and \emph{hannibal lecter}. In the last case, the connection arises because all the listed actors and characters appear in movies related to the cannibalistic serial killer Hannibal Lecter. Indeed, the most distinctive tag for this neuron is \emph{hannibal lecter}, arguably the most specific descriptor of this feature. 

To explore the limitations of neuron interpretability in certain edge cases, consider neuron \emph{id3125}, which emerges as the top neuron for the tags \emph{space action}, \emph{space adventure}, and \emph{classic sci-fi}. This overlap suggests limited granularity in one or more aspects: (1) the SAE may lack sufficient expressiveness to distinguish certain closely related concepts, potentially due to a small embedding width or neuron collapse during training; (2) the SAE training or labeling data -- both containing solely user \emph{interaction data} -- may not capture fine-grained semantic differences; or (3) the conceptual differences may be inherently vague or ambiguous. In this example, the most likely culprit is a combination of (3) and item popularity bias. Specifically, \emph{Star Wars: Episode IV - A New Hope (1977)}, the most frequently tagged item, accounts for over 0.6\% of all tag assignments and is associated with all of the above \emph{space} tags. This shared association entangles the corresponding tag representations at the item level, making them difficult to distinguish through our method, which aggregates tag--neuron links across items.

\subsection{The Effect of Steering on User Embeddings}
In addition to the results presented in Section \ref{sec:steeringEval}, we also explored the overall effect of the steering procedure on users' embeddings of \texttt{ELSA+L2} model. Figure~\ref{fig:steering_global} shows UMAP~\cite{mcinnes2018umap} projections of $1,000$ randomly sampled user embeddings (grey) and their steered reconstructions (in color, with increasing saturation) for $\alpha \in \{0.2, 0.4, 0.6, 0.8\}$ across three different concepts. 
The colored dots (e.g., red in the first plot) vary in intensity, with lighter tones representing smaller values of $\alpha$ (e.g., $\alpha = 0.2$) and darker tones indicating stronger steering (e.g., $\alpha = 0.8$).
In each of the three subplots, we highlight (using colored crosses) three movies most frequently tagged with the corresponding concepts.
As the influence increases, we observe convergence toward regions containing representative items: for example, many users move closer to \emph{Titanic} when boosting \emph{love story}, and toward \emph{Toy Story} when boosting \emph{children} (see the first and third plots in Figure~\ref{fig:steering_global}). Moreover, note the interesting pattern that emerges when boosting the neuron corresponding to the acclaimed director David Lynch (see middle of Figure~\ref{fig:steering_global}, where we highlight three of Lynch's movies): user embeddings converge toward \emph{Blue Velvet} (neo-noir thriller) and \emph{Mulholland Drive} (surreal psychological mystery), with comparatively less shift toward \emph{Eraserhead} (avant-garde horror). This contrast demonstrates that our approach provides individual personalization, unlike mechanisms that simply add concept-related items to the recommendation list or re-rank globally.

\end{document}